\documentclass[journal,twoside]{IEEEtran}
\usepackage[utf8]{inputenc}
\usepackage[T1]{fontenc}
\usepackage{cite}
\usepackage{amsmath,amssymb}
\usepackage{graphicx}
\usepackage{booktabs}
\usepackage{multirow}
\usepackage{siunitx}
\usepackage{xcolor}
\usepackage[hidelinks]{hyperref}
\usepackage{xurl}
\usepackage{orcidlink}

\graphicspath{{figures/}}

\newcommand{\Prx}{P_{\mathrm{rx}}}

\newcommand{\LunarRF}{LunarRF}
\definecolor{hrefblue}{RGB}{0,0,238}
\hypersetup{colorlinks=true, citecolor=hrefblue, linkcolor=hrefblue, urlcolor=hrefblue}

\begin{document}
\bstctlcite{IEEEexample:BSTcontrol}

\title{\LunarRF: A Terrain-Aware Propagation Engine and Pole-Wide 3GPP Surface Coverage Atlas of the Lunar South Pole}

\author{Christopher~J.~Santorelli\,\orcidlink{0000-0001-8666-7890}$^{1}$ and Alexander~M.~Wyglinski\,\orcidlink{0000-0002-3357-0064}$^{2}$\\[2pt]
{\normalsize $^{1}$Department of Computer Science, Worcester Polytechnic Institute, Worcester, MA 01609 USA}\\
{\normalsize $^{2}$Department of Electrical and Computer Engineering, Worcester Polytechnic Institute, Worcester, MA 01609 USA}%
\thanks{Corresponding author: C.~J.~Santorelli (\mbox{cjsantorelli@wpi.edu}).}}

\markboth{Preprint, 19 October 2026}%
{Santorelli and Wyglinski: LunarRF: A Terrain-Aware Propagation Engine and 3GPP Coverage Atlas of the Lunar South Pole}

\IEEEpubid{This work has been submitted to the IEEE for possible publication. Copyright may be transferred without notice, after which this version may no longer be accessible.}

\maketitle

\begin{abstract}
Artemis surface operations will depend on 3GPP cellular links, yet no public product shows where such links are physically possible across the lunar south pole. We present \LunarRF, a GPU system for terrain-aware surface coverage, and the pole-wide atlas produced with it. A footprint engine casts \num{8192} radial profiles from one of \num{11827} candidate masts enumerated by stated rules on the orbit-adjusted LOLA elevation model, takes the viewshed, and evaluates ITU-R P.1812 Delta-Bullington propagation on every profile, continuous through the radio horizon. An order-independent resolver merges the footprints per \SI{20}{\metre} cell, making the run shardable and resumable. The atlas covers the disc south of \ang{80}S at \SI{20}{\metre} for four SFCG-recommended surface bands and the IM-2 demonstration band, 1.8 to \SI{25.375}{\giga\hertz}, under a stated link budget, storing per cell the best-server power and identity, redundancy count, line of sight and coverability. The engine is validated against a parabolic-equation solver, ray tracing and analytic solutions. Swapping Deygout for Delta-Bullington, or adding terrain noise at the level of the published height-error map of the DEM, moves the coverable class on under \SI{1}{\percent} of a mast's disc in the mean. With every candidate active, \SI{90.5}{\percent} of the disc is coverable at \SI{1.8}{\giga\hertz}, \SI{59}{\percent} at \SI{5.855}{\giga\hertz} and \SI{10}{\percent} at \SI{25.375}{\giga\hertz}, or \SI{84}{}, \SI{43}{} and \SI{5}{\percent} after the measured \SIrange{3}{4}{dB} engine optimism. A mast covers about \SI{90}{\percent} of its visible area within its budget-set range, and no band is terrain-limited. The nine current Artemis regions are \SIrange{85}{99}{\percent} coverable at \SI{1.8}{\giga\hertz} and \SIrange{22}{85}{\percent} at \SI{5.855}{\giga\hertz}. The atlas and candidate table are released with a DOI.
\end{abstract}

\begin{IEEEkeywords}
3GPP, 5G mobile communication, Artemis, coverage planning, diffraction, digital elevation models, graphics processing units, lunar surface communications, radio propagation, site selection, terrain factors.
\end{IEEEkeywords}

\section{Introduction}
\label{sec:intro}

\IEEEPARstart{T}{he} Artemis campaign will place crews, rovers and infrastructure within a few degrees of the lunar south pole, where the terrain is among the most rugged on the Moon and the Sun never rises more than a few degrees above the horizon. This region is characterized by heavily cratered highland terrain, kilometer-scale reliefs, and steep slopes mantled in loose, fine-grained regolith with unique electromagnetic properties \cite{olhoeft1975}, while a maximum solar elevation of approximately \ang{1.5} \cite{mazarico2011} casts long shadows that have the potential to obscure surface hazards. Surface operations will rely on robust and resilient wireless communication links between the lander, the Lunar Terrain Vehicle (LTV), astronauts conducting extravehicular activity (EVA) and unattended instruments distributed across the landing region. The Space Frequency Coordination Group (SFCG) has recommended
specific frequency ranges for wireless systems deployed to the lunar
surface in Recommendation SFCG 32-2R5~\cite{sfcg32}, which map onto
3GPP operating bands supporting both Long-Term Evolution (LTE) and 5G
New Radio (NR)~\cite{edwards2023}. This atlas evaluates four of those
bands together with the 1.8\,GHz band of the IM-2
demonstration~\cite{edwards2023}. In March 2025, Nokia Bell Labs' Lunar Surface Communications System
(LSCS), the first LTE network on the Moon, reached Mons Mouton, one of
the candidate Artemis III landing regions, aboard Athena, the Nova-C
lander of Intuitive Machines' IM-2 mission~\cite{im2mission,nasaim2end}.
Although the network powered on and exchanged telemetry data with
Earth, the Athena lander came to rest on its
side~\cite{nasaim2end}. The rover and hopper it carried could not be
deployed and no connection to the network was
established~\cite{im2mission,nasaim2end}. Link-budget analysis for lunar surface networks requires the path loss between every transmitter and every location a receiver may occupy. As the Moon lacks an atmosphere, this path loss is governed almost entirely by free-space spreading, ground reflection, and terrain obstruction.

\IEEEpubidadjcol
The lunar terrain constrains wireless propagation through both the proximity of the horizon and the scale of its relief. Over a smooth sphere of radius \SI{1737.4}{\kilo\metre}, a \SI{4}{\metre} mast and a \SI{1.5}{\metre} receiver have a combined radio horizon of only \SI{6.0}{\kilo\metre} (\SI{3.7}{\kilo\metre} plus \SI{2.3}{\kilo\metre} respectively), beyond which coverage depends on the elevation of the mast relative to the surrounding terrain. Crater rims and massifs further obstruct line of sight (LOS) over distances of tens of kilometers, making local topography, rather than transmitter range, the governing factor in coverage. The permanently shadowed regions (PSRs), which host the volatile deposits motivating polar exploration, illustrate this constraint in a counterintuitive way. Terrain that shields a crater floor from sunlight arriving at elevations below approximately \ang{1.5} does not necessarily block a mast on the overlooking rim, since the mast views the floor from above rather than across the horizon. Consequently, \SI{61}{\percent} of the PSR area within the study region is in LOS of at least one permitted site, and \SI{59}{\percent} is reachable at \SI{1.8}{\giga\hertz} (\SI{53}{\percent} at \SI{2.58}{\giga\hertz}, Table~\ref{tab:psr}). The terrain a smaller mast cannot serve lies instead on the shadowed inner flanks of craters and in small craters with no candidate site in view, and \SI{69}{\percent} of this unserved area at \SI{1.8}{\giga\hertz} lies outside permanent shadow (Section~\ref{sec:res_disc}). Surface network coverage is therefore a geographic problem that must be resolved with a digital elevation model (DEM) prior to network design.

However, there is no publicly available dataset or product which quantifies achievable network coverage across the polar region as a whole. Existing lunar propagation studies are limited to one or a few sites, with examples such as ray-traced path-loss modeling along a single crater rim, DEM-based coverage estimates for a small number of candidate sites, and deployment analyses which involve a few towers (Section~\ref{sec:related}). This paper addresses that gap with a \emph{potential-coverage atlas} of the lunar south polar region. Rather than optimizing mast placement, we enumerate every plausible mast site south of \ang{80}S, compute the coverage footprint of each with a terrain-aware propagation engine, and merge the footprints into pole-wide layers. For every \SI{20}{\metre} cell, these layers indicate whether any candidate mast can serve it, the achievable link quality, the sites from which it is served, and the number of alternative sites available. The design of the atlas serves two purposes. Prior to a mission it provides planners with a screening product for assessing the coverage potential of candidate regions and an input for subsequent network-design studies. After landing, the data contained in this study can support an onboard or crew-operated tool that identifies, from the actual landing location, the nearest candidate mast site offering the greatest coverage.

The principal contributions of this paper are summarized as follows:
\begin{enumerate}
\item A GPU atlas engine that combines a radial viewshed with ITU-R P.1812 Delta-Bullington propagation on every receiver, carrying the grazing ground reflection through its spherical-Earth term, with a link budget derived from stated noise and power assumptions in both link directions (Section~\ref{sec:engine}).
\item A candidate-site methodology with two classes, terrain high points anywhere on the disc and landable \SI{1}{\kilo\metre} grid cells inside the Artemis candidate landing regions, that produces \num{11827} sites from stated rules with no hand selection (Section~\ref{sec:candidates}).
\item The atlas itself, with continuous best-server received power, best-server identity, redundancy count and coverable mask at \SI{20}{\metre} for four SFCG-recommended surface bands and the IM-2 band, computed for all \num{11827} sites on the WPI Turing cluster (Section~\ref{sec:computation}), with a compact \SI{160}{\metre} release product and per-region \SI{20}{\metre} windows.
\item Validation against a parabolic-equation solver on real terrain profiles and against NVIDIA Sionna~RT ray tracing on canonical obstacles and real terrain, including the limitation of wedge-based first-order diffraction on DEM meshes, and quantified sensitivity of the coverage class to the diffraction model, the DEM product, DEM error and candidate density across latitude bands (Section~\ref{sec:validation}).
\item Pole-wide and per-region results for the nine current Artemis candidate landing regions and three regions of the 2022 list, showing that within its free-space range a mast covers about \SI{90}{\percent} of its visible area at every band, that the range shrinks from \SI{22}{\kilo\metre} to \SI{1.6}{\kilo\metre} across the bands while a \SI{20}{dB} stronger budget reaches visible terrain \SI{100}{\kilo\metre} away, that most of the permanently shadowed area is visible from the permitted rim sites, and that coverage is effectively single-server at \SI{25.375}{\giga\hertz} (Section~\ref{sec:results}).
\item The finding that local-maximum site selection on the PDS LDEM\_80S\_20M product picks orbit-track artifacts of \SIrange{20}{257}{\metre} at about one site in ten. The atlas uses the orbit-adjusted PGDA product, on which no selected site is more than \SI{20}{\metre} higher in the PDS product, and carries the height difference between the two products for every site.
\end{enumerate}

\begin{figure*}[t]
\centering
\includegraphics[width=0.9\textwidth]{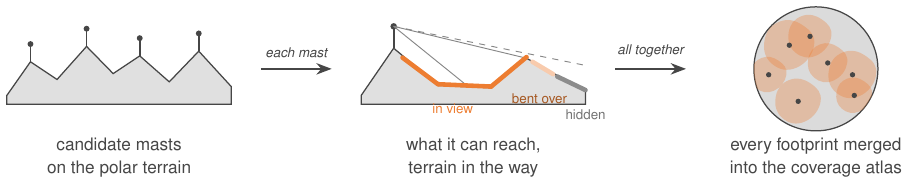}
\caption{Candidate \LunarRF\ mast sites are enumerated on the
south-polar terrain (left). For each mast, the system computes what it
can reach with the terrain in the way, the ground in view, the ground
just behind a rim that diffraction still reaches, and the ground that
is hidden (center). Every footprint is merged into one pole-wide
coverage atlas, darker where several masts overlap (right).
Fig.~\ref{fig:level0} names the inputs, the processing loop and the
outputs of the system outlined here, while Fig.~\ref{fig:level1}
details the operations inside the processing block.}
\label{fig:system_glance}
\end{figure*}
Four products, namely the full-resolution atlas, the compact
\SI{160}{\metre} release product, the per-region \SI{20}{\metre} windows,
and the \num{11827}-site candidate table, are publicly released under a
\href{https://doi.org/10.5281/zenodo.23301640}{DOI}.

The remainder of this paper is organized as follows.
Section~\ref{sec:related} reviews lunar surface propagation models,
terrain-based lunar visibility products, ray tracing and terrestrial
coverage prediction, and places the atlas among them.
Section~\ref{sec:architecture} describes the system in the order in which
it runs. It covers the terrain, shadow, region and regolith inputs
(Section~\ref{sec:data}), the two classes of candidate mast sites and the
rules that produce them (Section~\ref{sec:candidates}), and the footprint
engine with its link budget, viewshed and Delta-Bullington propagation
(Section~\ref{sec:engine}). It closes with the computation
(Section~\ref{sec:computation}), in which the tiling and resolver merge
\num{11827} footprints into one pole-wide atlas on the cluster and the
exact statistics behind every reported figure are defined.
Section~\ref{sec:results} opens with the validation of the engine against
a parabolic-equation reference, ray tracing and analytic solutions and the
sensitivity of the coverage class to the model and the DEM
(Section~\ref{sec:validation}). It then presents the pole-wide, per-mast
and per-region results, together with the throughput, power-control and
reference-budget studies, and closes with a discussion of the scope and
limitations of the atlas (Section~\ref{sec:discussion}).
Section~\ref{sec:conclusion} concludes and outlines future work.     %
Fig.~\ref{fig:system_glance} previews the system at its simplest. Candidate masts are enumerated on the polar terrain, the footprint of each is computed against that terrain, and all footprints are merged into one pole-wide atlas. Section~\ref{sec:architecture} defines it exactly.
\section{Background and Related Work}
\label{sec:related}

\subsection{Lunar surface propagation models}
Early lunar channel studies adapted existing terrestrial models. Foore and Ida introduced a modified Longley--Rice Irregular Terrain Model (ITM) \cite{hufford1982}, adapted for the lunar radius and the absence of an atmosphere, to predict path loss over the lunar surface \cite{foore2007}. Hwu \emph{et al.}\ studied lunar terrain effects on surface communications using computational electromagnetics, taking the dielectric characteristics of the regolith as input alongside terrain geometry and antenna placement \cite{hwu2008}. Chizhik \emph{et al.}\ developed the Analytic Propagation Approximation over variable terrain (APA), which approximates the intervening terrain by a constrained parabola and computes signal strength via a modal (Airy function) solution. Intended for rapid lunar coverage prediction, APA was validated
against over 3000 measured links in the desert terrain of
Fuerteventura, Spain, achieving 8.5\,dB RMS
error~\cite{chizhik2020}. The campaign itself was a Nokia Bell Labs
lunar-analog measurement at 1.8\,GHz with a 4\,m transmit antenna,
the band and mast height of the later IM-2
demonstration~\cite{chizhik2020,edwards2023}. Edwards \emph{et al.}\ reported a NASA study in which LOLA topography of a \SI{27.5}{\kilo\metre} square around the Shackleton connecting ridge was imported at \SI{55}{\metre} resolution into Remcom's Wireless InSite ray tracer \cite{edwards2023}. Configured with lunar curvature, no atmosphere, and regolith electrical properties, the simulation mapped path loss and throughput for a \SI{30}{\metre} lander-mounted base station and a \SI{3.5}{\metre} LTV user. Edwards \emph{et al.}\ also summarized the SFCG-recommended 3GPP bands and the Artemis EVA range limits (\SI{2}{\kilo\metre} walking, \SI{10}{\kilo\metre} LTV-assisted), which this paper adopts along with the study's representative link parameters. Wang and Wyglinski later proposed the Barren, Irregular, Chaotic Terrain Ring (BICTR) model, a stochastic ring-of-reflectors multipath channel model inspired by Jakes' and existing two-ray models \cite{wang2025bictr}. BICTR was validated against the NASA Desert RATS 2022 lunar-analog measurement campaign, where it predicted received power on average \SI{7.52}{dB} and \SI{9.90}{dB} closer to measurement than ITM \cite{hufford1982} and ITWOM \cite{shumate2010}, respectively \cite{wang2025bictr}. Although BICTR captures multipath and fading, it omits diffraction, the mechanism that dominates path loss behind terrain obstructions. Adebowale and Ostermann published a ray-traced path-loss model at \SI{5.855}{\giga\hertz} for Shoemaker Rim~F \cite{shoemakerrim}, and Pabari \emph{et al.}\ modeled radio-frequency propagation for a wireless sensor network on Chandrayaan-1 Terrain Mapping Camera DEMs of four sample sites \cite{pabari}. Qin \emph{et al.}\ searched the illuminated south-polar region for a main communication tower and placed sub-stations with a genetic algorithm to maximize field strength within a \SI{10}{\kilo\metre} radius, using a P.526-based terrain model \cite{qin2024}. These three works are the closest antecedents of the present work. Lordos \emph{et al.}\ proposed autonomously deployable towers for communication into permanently shadowed regions \cite{lordos2020}, a concept that bears directly on the PSR result of Section~\ref{sec:res_disc}. Our work quantifies what small masts on the illuminated uplands cannot reach, and such infrastructure represents one approach to extending coverage into these regions. These studies established the governing propagation physics, operational requirements, and site-specific coverage at one or a few locations. No published study provides coverage analysis at the scale of the full polar region, nor a systematic enumeration of candidate deployment sites. The present work addresses both, contributing a pole-wide coverage atlas in which candidate sites are enumerated and ranked.

\subsection{Terrain-based lunar visibility products}
Mazarico \emph{et al.}\ published pole-wide illumination and Earth-visibility maps computed by horizon analysis of the LOLA DEM \cite{mazarico2011}, and Barker \emph{et al.}\ released improved, orbit-adjusted south-polar elevation products with error estimates \cite{barker2021,barker2023}. Peña-Asensio \emph{et al.}\ combined illumination, Earth visibility, PSR proximity, and geological criteria in a multi-criteria assessment of \num{1247} candidate locations across the thirteen 2022 Artemis~III regions \cite{penaasensio2025}. Whereas these products address whether a given point can view the Sun or the Earth, the present work applies the same terrain data to visibility of a surface base station, and incorporates the propagation physics that a line-of-sight test alone omits.

\subsection{Ray tracing}
Differentiable GPU ray tracers such as NVIDIA Sionna RT~\cite{hoydis2023sionnart}, the ray-tracing module of the Sionna library~\cite{hoydis2022sionna}, offer path enumeration (reflection, first-order UTD wedge diffraction~\cite{kouyoumjian1974}, scattering) on triangle meshes. We use Sionna~RT as a cross-check of the atlas engine on line of sight and reflection (Section~\ref{sec:validation}); it is not a diffraction reference for gentle terrain, because wedge diffraction on a faceted mesh of a smooth crest returns little or no energy, which is a limitation of wedge-based diffraction on DEM meshes rather than of the software. Parabolic-equation (PE) methods \cite{levy2000,barrios1994,dockery1996} are the standard numerical reference for propagation over irregular terrain with lossy ground, and serve that role in our work.

\subsection{Terrestrial coverage prediction}
The atlas engine follows the structure of terrestrial planning tools, combining a viewshed with a point-to-area propagation model in which a spherical-Earth term carries the grazing ground reflection on clear paths and a knife-edge construction carries the diffraction on obstructed ones. The model is adapted to the Moon through the lunar radius, the absence of an atmosphere, regolith electrical properties, and a sphere-drop term applied to every profile. The Delta-Bullington model of ITU-R P.1812 \cite{itu1812} is the standard point-to-area diffraction method for terrestrial prediction up to \SI{6}{\giga\hertz}. Its application at \SI{25.375}{\giga\hertz} is an extrapolation beyond the recommendation's stated \SI{6}{\giga\hertz} limit. However, the components of the method retained here, namely the knife-edge diffraction integral and the spherical-surface term, are frequency-scaled physical constructions rather than empirical fits. Moreover, the terrestrial effects that motivate the stated range, namely clutter and atmospheric attenuation, are absent on the Moon. The Deygout \cite{deygout1966}, Bullington, and single knife-edge methods \cite{itu526} belong to the same family of knife-edge diffraction models and are retained as alternatives, allowing the sensitivity of the coverage classification to the choice of method to be quantified.

\section{System Architecture}
\label{sec:architecture}
\begin{figure*}[t]
\centering
\includegraphics[width=\textwidth]{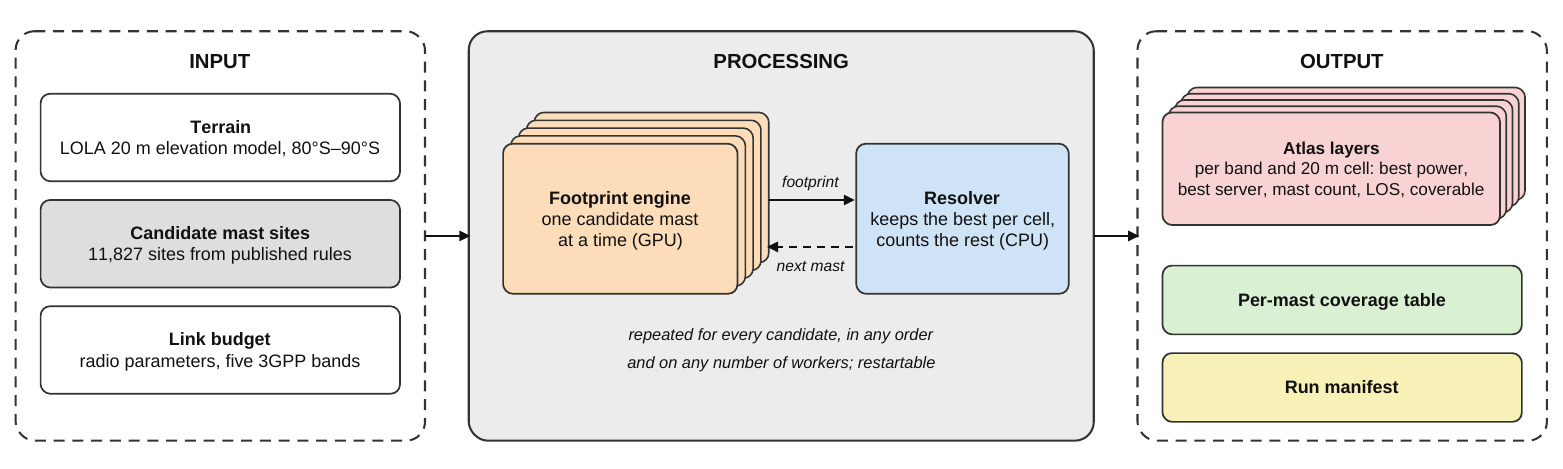}
\caption{The terrain, the candidate mast table and the link budget
enter the level-0 \LunarRF\ processing block as its three inputs. The
block produces the atlas layers, the per-mast coverage table and the
run manifest. Processing is a loop between two components with fully
separate state. The footprint engine owns one mast at a time on the
GPU, and the resolver owns the pole-wide accumulators on the CPU. The
footprint is the only data that crosses between them. The resolver
applies order-independent per-cell rules, so the candidate list can be
sharded and run in parallel, and a run that stops continues from its
manifest rather than from the start.}
\label{fig:level0}
\end{figure*}

This section presents the \LunarRF\ system architecture in the execution order of Fig.~\ref{fig:level0}. Section~\ref{sec:data} describes the input datasets, Section~\ref{sec:candidates} the frozen table of candidate mast sites that drives the processing loop, Section~\ref{sec:engine} the per-mast footprint engine, and Section~\ref{sec:computation} the resolver and run-management layer through which \num{11827} footprints are consolidated into a single pole-wide atlas. The design separates per-mast computation from pole-wide accumulation so that the loop can be sharded, interrupted, and resumed without coordination. Fig.~\ref{fig:level1} expands the processing block of Fig.~\ref{fig:level0} into the operations of the engine and the resolver. The parameter values behind each operation are given in the text of this section, and the field names of the released files in the data release. The validation studies of Section~\ref{sec:validation} read the same profiles and accumulators but never write to the store.

\subsection{Inputs: Terrain, Shadow, Regions and Regolith}
\label{sec:data}

All inputs are obtained from public products. The exact URLs, sizes and SHA-256 checksums are recorded in the provenance file of the data release. Each raster is handled in the LOLA south polar stereographic frame, a sphere of radius $R=\SI{1737.4}{\kilo\metre}$, scale $k=1$ at the pole, central meridian \ang{0}E along $+y$ and \ang{90}E along $+x$ (PROJ string \texttt{+proj=stere +lat\_0=-90 +lon\_0=0 +k=1 +R=1737400}). The map scale factor
\begin{equation}
k(\varphi)=\frac{2}{1+\sin|\varphi|}
\label{eq:scale}
\end{equation}
reaches \num{1.0077} at \ang{80}S (\SI{0.8}{\percent}) and is neglected in distances.
\\
\subsubsection{Elevation}
\label{sec:dem}
The atlas terrain is the orbit-adjusted LOLA product \texttt{LDEM\_80S\_20MPP\_ADJ} from the NASA Planetary Geodynamics Data Archive (PGDA) \cite{barker2023}: $\num{30400}\times\num{30400}$ cells at \SI{20}{\metre}, spanning $\pm\SI{304}{\kilo\metre}$, heights in meters above the reference sphere. The same grid and transform are used by the PDS Geosciences Node product \texttt{LDEM\_80S\_20M} V2.1 (float32 kilometers relative to $R$) \cite{neumann2010gdr}, which we keep as a comparison DEM for error propagation. The two products differ in a way that matters for site selection (Section~\ref{sec:candidates}), the PDS product carries smooth positive bumps of \SIrange{20}{257}{\metre} height and more than \SI{140}{\metre} width along orbit tracks at about one in ten of its local maxima, the adjusted product does not. Each candidate site therefore carries $\Delta z_{\mathrm{PDS}}$, its height in the PDS product minus its height in the PGDA product, as a per-site measure of inter-product elevation disagreement.

For validation at higher resolution we use the PGDA \SI{5}{\metre} LOLA mosaic of \ang{87}S--\ang{90}S \cite{barker2021}. There is no public \SI{5}{\metre} LOLA product that exists north of \ang{87}S.

\subsubsection{Permanent shadow}
Permanently shadowed regions are taken from the PGDA \SI{20}{\metre} product \texttt{LPSR\_80S\_20MPP\_ADJ} \cite{barker2023} on the atlas grid, with the LOLA EXTRAS average-visibility and PSR maps at \SI{120}{\metre} and \SI{60}{\metre} \cite{mazarico2011,neumann2010gdr} used for figures. Masts are not sited in permanent shadow regions as a solar-powered lander or mast would not be placed there. PSR floors remain valid receiver locations and are reported as coverable when they have line of sight and margin because a rover or astronaut may operate there.

\subsubsection{Landing regions}
\label{sec:regions}
Two NASA lists define the regions of interest. The release of 19~August~2022 named thirteen candidate regions of roughly $\SI{15}{\kilo\metre}\times\SI{15}{\kilo\metre}$ \cite{nasa2022}. The update of 28~October~2024 named nine, of which five were kept from 2022 and four added \cite{nasa2024}. NASA's February 2026 architecture update makes Artemis~III a low-Earth-orbit rendezvous test and moves the first crewed landing to Artemis~IV \cite{nasa2026}. The nine regions were defined for Artemis~III \cite{george2026} and remain the only published reference set and we refer to them as the \emph{Artemis candidate landing regions}. Region footprints are the extents of the \SI{1}{\kilo\metre} Lunar Grid Reference System navigational grids published by the USGS for the nine 2024 regions \cite{mcclernan2024a} and, where public, for the 2022 regions \cite{mcclernan2024b}. The region polygons are reprojected from the USGS polar stereographic frame (false easting and northing \SI{500}{\kilo\metre}, scale factor \num{0.994}) into the LOLA frame. Twelve regions have a public footprint, the nine current regions plus Faustini Rim~A, Connecting Ridge and Amundsen Rim from the 2022 list. The five 2022 regions without a public grid (Peak near Shackleton, Connecting Ridge Extension, de~Gerlache Rim~1, de~Gerlache--Kocher Massif, Leibnitz Beta Plateau) are kept in the configuration as unverified Gazetteer centers and are excluded from all figures and tables. The polygons of Pe\~na-Asensio \emph{et al.}\ \cite{penaasensio2025} are the closest citable replacement and could be adopted in a future run.

\subsubsection{Regolith}
The ground enters the propagation model through its admittance in the spherical-Earth term of Delta-Bullington (Section~\ref{sec:tworay}), with a relative permittivity $\varepsilon_r=3.7$ and loss tangent $\tan\delta=0.01$, within the range of the Apollo-sample measurements of Olhoeft and Strangway \cite{olhoeft1975} for compacted regolith of density \SI{\approx 2}{\gram\per\cubic\centi\metre}; $\tan\delta=0.1$ is retained as a high-loss sensitivity case. Conductivity follows as
\begin{equation}
\sigma=2\pi f\,\varepsilon_0\,\varepsilon_r\tan\delta .
\label{eq:sigma}
\end{equation} No small-scale roughness term is applied, although sub-centimeter and sub-pixel roughness of the regolith is well characterized~\cite{helfenstein1999,bandfield2015}. The same values define the impedance ground of the parabolic-equation reference (Section~\ref{sec:val_pe}), and roughness below the DEM resolution is among the omissions listed in Section~\ref{sec:discussion}.

\begin{figure*}[t]
\centering
\includegraphics[width=\textwidth]{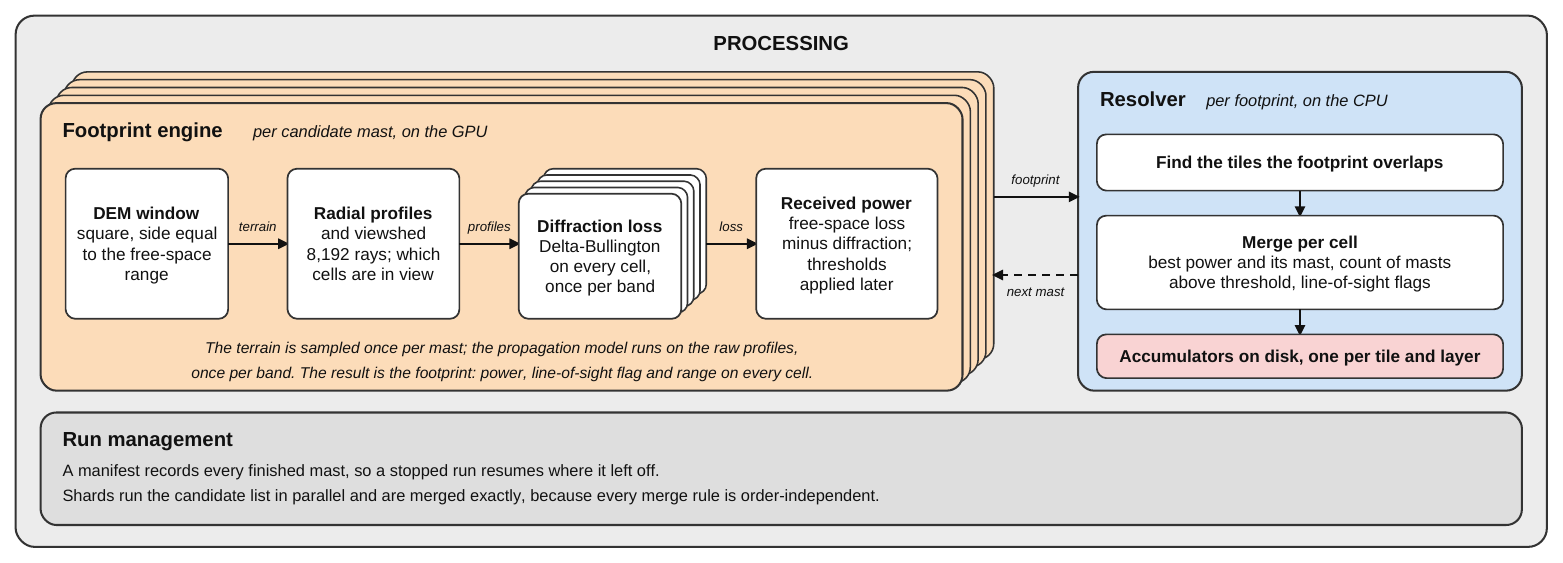}
\caption{The engine evaluates the terrain geometry once per mast on
\num{8192} radial profiles. It evaluates the propagation loss once per
band on those same profiles. The loss model therefore sees the raw,
unsmoothed terrain. The footprint carries the received power, the
line-of-sight flag and the range of every cell. Nothing is thresholded
or cached inside the engine. The resolver merges each footprint into
the per-tile accumulators without knowing which mast it is merging,
which shard produced it, or in what order the footprints arrive. This
independence allows the run to be sharded and resumed. Run management
is shared by both halves. It is the only place that reads the
manifest. The link budget, the diffraction model and the merge rules
are given in Sections~\ref{sec:engine} and \ref{sec:computation}.}
\label{fig:level1}
\end{figure*}

\subsection{Candidate Mast Sites}
\label{sec:candidates}

The atlas does not choose where masts go. It enumerates every place a mast could plausibly stand and computes the coverage of each. The union over candidates is the coverable map, the per-cell count is a measure of serving redundancy, and the per-candidate table lets a planner or an optimiser select a subset. Two classes of place are enumerated by rules that are all stated parameters (listed with the released run parameters).

\subsubsection{Rules common to both classes}
Sites are extracted per \SI{40}{\kilo\metre} tile with a \SI{10}{\kilo\metre} halo (Section~\ref{sec:computation}) such that every site sees its full neighborhood and no site is counted twice. Sites north of \ang{80}S (DEM corners) and inside permanent shadow are dropped. Within a class, sites are thinned to a minimum spacing of \SI{1}{\kilo\metre} by a greedy pass that keeps the highest-scoring site first, \SI{1}{\kilo\metre} is the scale at which coverage is assessed. The spacing does not limit the density: the local-maximum, prominence, permanent-shadow and slope rules have already fixed which cells qualify, and re-extracting the whole pole with the same rules at \SI{500}{\metre} admits only \SI{7}{\percent} more sites, \num{12683}. The atlas itself was run at \SI{500}{\metre} spacing on three test tiles (Section~\ref{sec:val_convergence}). Sites within the ray radius (\SI{22}{\kilo\metre}) of a verified region footprint are processed first (\texttt{priority~0}, \num{4020} sites) such that a partial run already answers the landing-region questions.

\subsubsection{Ridge class: where the terrain offers a high point}
A cell qualifies if it is either (i) a \emph{local maximum}, the highest cell of its $\SI{1}{\kilo\metre}\times\SI{1}{\kilo\metre}$ neighborhood, standing at least \SI{50}{\metre} above the lowest cell of that neighborhood (prominence), or (ii) a \emph{ridge or rim cell}: Laplacian mean curvature more negative than \SI{-0.006}{\per\metre} after a two-pixel Gaussian smooth (a convex-up crest), thinned to one cell wide; and in either case its local slope at \SI{20}{\metre} is at most \ang{12}, a mobility limit for a small lander or a rover-placed tripod rather than a landing limit. Surviving cells are ranked by height (with a \SI{50}{\metre} bonus for true peaks), thinned to \SI{1}{\kilo\metre} and capped at 150 per tile (the cap binds on no tile: the largest count is 146, the median 46). The result is \num{10499} ridge-class sites, \num{7119} of them in the \ang{80}--\ang{83}S band, which holds the most area and the densest field of local maxima. Rhe polar band is the more rugged in the sense that matters for coverage (larger relief and PSR area per unit area, Section~\ref{sec:res_disc}). This class is deliberately not a landability screen: rims and peaks are often too steep for a lander at the \SI{100}{\metre} scale. It answers the question ``if a \SI{4}{\metre} mast could be put on the best terrain, what could it see?''

\subsubsection{Landing-site class: where a lander can stand}
Sites of this class lie only inside the twelve verified USGS region footprints. For each region we take the centers of the \SI{1}{\kilo\metre} grid cells aligned to the footprint (the USGS grid itself), keep a cell if its \SI{100}{\metre}-scale slope is at most \ang{5}, the flatness requirement of the Human Landing System \cite{penaasensio2025}, and drop cells in permanent shadow. The slope is computed on the DEM after smoothing with a \SI{100}{\metre} box filter. No further thinning is applied. The result is \num{1328} sites. Mons Mouton Plateau, by far the largest region, contributes \num{766}, and Peak near Cabeus~B, the steepest, contributes four (Table~\ref{tab:cand_regions}). The landable fraction of the grid cells ranges from \SI{25}{\percent} at Slater Plain to \SI{1}{\percent} at Peak near Cabeus~B. In the base atlas both classes carry a \SI{4}{\metre} mast, the height at which the IM-2 base station was carried on an upper panel of the Nova-C lander \cite{im2install}, so that the layers are directly comparable. Taller landing-site masts of \SI{10}{\metre} and \SI{30}{\metre} represent lander-mounted equipment on larger vehicles such as Blue Moon Mk~2 (about \SI{16}{\metre} tall) and Starship HLS (\SI{52}{\metre} tall) \cite{creech2025hls}, whose antenna mounting heights are unpublished, they form separate sensitivity sets, of which the \SI{30}{\metre} case is the reference scenario of Section~\ref{sec:res_reference} and the others are future work.

\begin{table*}[!t]
\caption{Candidate sites per region (12 verified footprints; numbers as in Fig.~\ref{fig:candidates_pole}). Footprints are the USGS navigational-grid extents, not NASA region polygons (hence the \SI{4579}{\kilo\metre\squared} of Mons Mouton Plateau). Landing-site class inside the footprint, ridge class inside it, and ridge class within \SI{8}{\kilo\metre} of it. Mons Mouton overlaps Mons Mouton Plateau (\SI{138}{\kilo\metre\squared}) and Nobile Rim~1 (\SI{89}{\kilo\metre\squared}), and Nobile Rim~1 overlaps the Plateau (\SI{43}{\kilo\metre\squared}). Per-region counts are not additive, the last row gives the unique totals. Bold marks the highest value in each column and italic the lowest.}
\label{tab:cand_regions}
\centering\footnotesize
\begin{tabular}{@{}rllrrrr@{}}
\toprule
No. & Region & List & Area (km$^2$) & Landing-site & Ridge inside & Ridge $\leq 8$ km \\
\midrule
1 & de Gerlache Rim 2 & 2024 & 446 & 44 & 8 & 43 \\
2 & Haworth & 2024 & 941 & 137 & 14 & 23 \\
3 & Malapert Massif & 2024 & 468 & 21 & 9 & 16 \\
4 & Mons Mouton & 2024 & 293 & 53 & 5 & 24 \\
5 & Nobile Rim 1 & 2024 & 446 & 53 & 5 & 19 \\
6 & Nobile Rim 2 & 2024 & 425 & 65 & 16 & 36 \\
7 & Peak near Cabeus B & 2024 & 446 & \textit{4} & \textit{3} & \textit{11} \\
8 & Slater Plain & 2024 & 446 & 107 & 14 & 52 \\
9 & Mons Mouton Plateau & 2024 & 4,579 & \textbf{766} & \textbf{138} & \textbf{171} \\
10 & Faustini Rim A & 2022 only & 425 & 69 & 13 & 38 \\
11 & Connecting Ridge & 2022 only & 293 & 26 & 9 & 15 \\
12 & Amundsen Rim & 2022 only & 425 & 50 & 11 & 30 \\
\midrule
 & All candidates (unique) & & & 1328 & 10499 & \\
\bottomrule
\end{tabular}

\end{table*}

\begin{figure*}[!t]
\centering
\includegraphics[width=\textwidth]{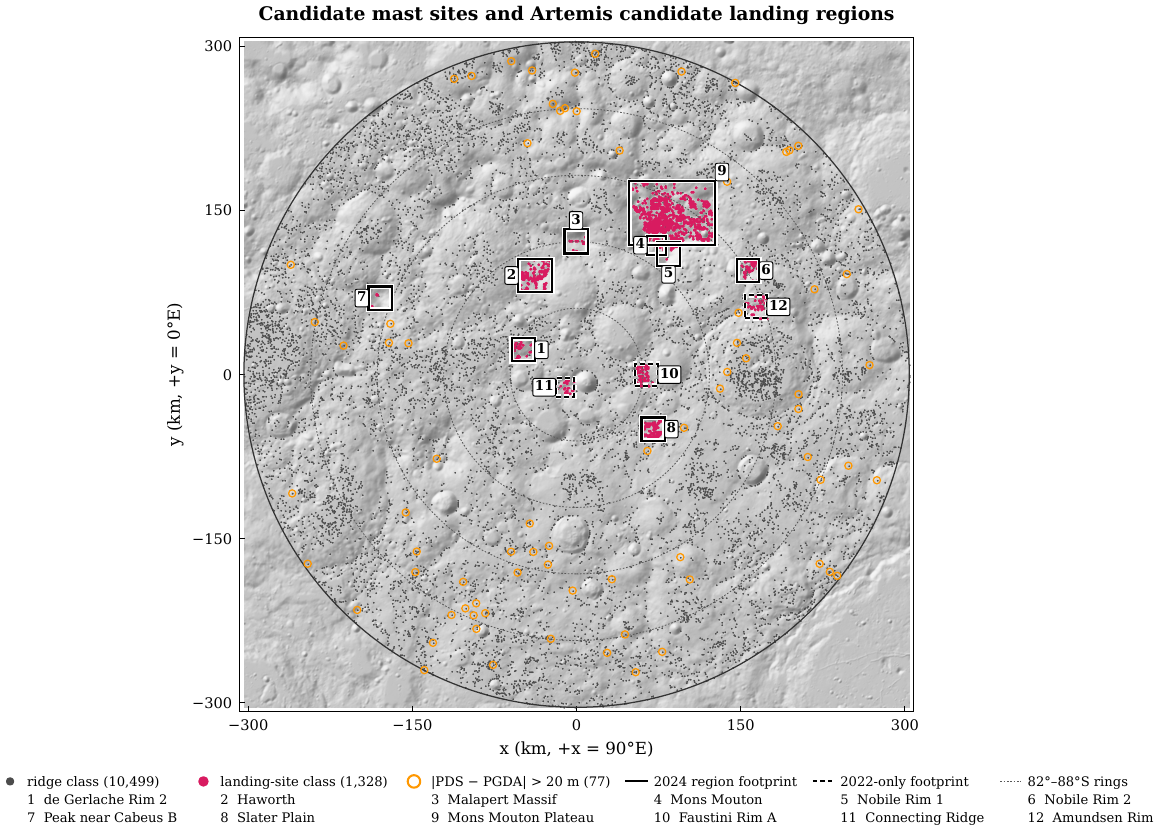}
\caption{Candidate mast sites are shown on the orbit-adjusted LOLA
\SI{20}{\metre} DEM in south polar stereographic projection,
\ang{80}S--\ang{90}S. Gray marks the \num{10499} ridge-class sites, and
magenta marks the \num{1328} landing-site-class sites. Orange rings
mark the ridge sites whose height differs by more than \SI{20}{\metre}
between the PDS LDEM\_80S\_20M and the adjusted product ($n=77$). All
77 are higher in the adjusted product (Section~\ref{sec:dem_artefacts}).
Black outlines show the nine 2024 region footprints as solid lines and
the 2022-only footprints as dashed lines, numbered beside each
footprint.}
\label{fig:candidates_pole}
\end{figure*}

\subsubsection{Criteria deliberately not used}
The candidate table is defined by terrain alone, and no further selection should be inferred from it. Illumination and solar power at the site are intentionally excluded as a candidate in a poorly lit location remains a valid coverage question, and the LOLA average-visibility maps allow that coupling to be introduced at a later stage. Earth visibility for direct-to-Earth backhaul is outside the scope of a surface-to-surface atlas. The distance from a ridge site to any landing region, and the geological, resource or scientific value of a site, are planning considerations rather than propagation ones and belong to a subsequent site-selection step applied to the released table. Overlapping footprints are neither optimized nor de-duplicated, every candidate transmits in the envelope, and the selection of a minimal set is future work. Finally, regolith bearing strength, boulders and roughness below the \SI{20}{\metre} DEM resolution cannot be assessed from the available terrain data and are omitted.

\subsubsection{DEM artifacts and site selection}
\label{sec:dem_artefacts}
The two \SI{20}{\metre} products derive from the same LOLA shots and differ only by the orbit adjustment. Their difference is not a terrain-error estimate (Section~\ref{sec:validation}). However, it does reveal a selection effect of the candidate rules. The local-maximum detector preferentially selects cells whose height is raised by an artifact. The artifacts are concentrated in the candidate set far beyond their share of the terrain. When the candidate rules of Section~\ref{sec:candidates} were first applied to the PDS product (\num{13415} sites), the site-height difference $\Delta z_{\mathrm{PDS}}$, PDS minus PGDA, was below \SI{1}{\metre} RMS for the landing-site class but \SI{16.5}{\metre} RMS for the ridge class and one-sided: \SI{10.0}{\percent} of ridge candidates (\SI{8.6}{\percent} in the \ang{80}--\ang{83}S band, \SI{14.9}{\percent} in \ang{83}--\ang{86}S and \SI{13.7}{\percent} in \ang{86}--\ang{90}S) sat on features \SIrange{20}{257}{\metre} higher in the PDS product than in the adjusted product. Inspection of the largest cases shows smooth positive bumps more than \SI{140}{\metre} across in the PDS DEM that are absent from the adjusted DEM; orbit-track errors rather than single-pixel spikes, which the detector selects precisely because they are local maxima. A mast placed on such a feature overstates its footprint. The atlas is therefore run on the adjusted product. In the final candidate set, detected on that product, no site is more than \SI{20}{\metre} higher in the PDS DEM, so the artifacts are no longer selected; the ridge-class RMS difference falls to \SI{3.8}{\metre}, and 77 sites (\SI{0.7}{\percent}) are more than \SI{20}{\metre} higher in the adjusted product than in the PDS product (Fig.~\ref{fig:candidates_pole}). Every site carries $\Delta z_{\mathrm{PDS}}$. The orbit-adjusted product exists because such errors were known; the point here is the degree to which peak-picking on the unadjusted product concentrates them. Cells where the two products differ by more than \SI{20}{\metre} occupy \SI{0.05}{\percent} of the pixels, and those where the PDS product is the higher \SI{0.02}{\percent}, yet they carry one ridge site in ten, a concentration of several hundred times. The 77 adjusted-product-higher sites are inspected individually (hillshade, profiles, LOLA shot count and local-maximum status in each product) in the candidate-freeze record of the release. All 77 are kept in the run and 60 are flagged in the released table, because only 23 are \SI{1}{\kilo\metre} local maxima in the adjusted product (none is in the PDS product) and 28 stand on cells with no LOLA shot, so that their heights rest on interpolation in both products. A companion test compares the LOLA shot count at candidate cells with that of all cells in the same latitude band, per detector, to check that neither the local-maximum nor the curvature rule prefers interpolated cells (Section~\ref{sec:val_effres}).
\subsection{Propagation Engine and Link Budget}
\label{sec:engine}

\subsubsection{Link budget and thresholds}
\label{sec:budget}
The base atlas uses a modest small-lander budget. A base-station can transmit power \SI{30}{dBm} and user-equipment (UE) power \SI{23}{dBm} (power class~3 \cite{3gpp36101}), isotropic (\SI{0}{dBi}) vertically polarized antennas at both ends, a \SI{4}{\metre} mast and a \SI{1.5}{\metre} receiver, an astronaut or rover-mounted antenna. Isotropic gains are used because the atlas is a site-potential envelope: a sector pattern implies a pointing decision and belongs to a deployment study. Noise is
\begin{equation}
N=k_{\mathrm{B}}\,\bigl(T_{\mathrm{ant}}+T_0(F-1)\bigr)\,B
\label{eq:noise}
\end{equation}
with a downlink carrier of $B = \SI{20}{\mega\hertz}$, an assumed UE
noise figure $F = \SI{7}{\decibel}$ and an assumed base-station noise
figure of \SI{3}{\decibel}, both better than the minimum receivers
implied by the reference-sensitivity requirements
of~\cite{3gpp36101}, \cite{3gpp36104}, and an antenna temperature
$T_\mathrm{ant} = \SI{150}{\kelvin}$, an assumed value between the cold
sky and the regolith brightness temperature. Polar regolith brightness temperatures are lower than equatorial values \cite{wei2019,morabito2006}, so \SI{150}{\kelvin} overstates the noise rather than understating it; a \SI{100}{\kelvin} change moves the thresholds by about \SI{1}{dB}. The required SNR is \SI{0}{dB} over the full \SI{20}{\mega\hertz} carrier with no fading, implementation or body-loss margin, i.e.\ a detection edge rather than a service threshold because the stored layer is continuous received power, any stricter requirement can be applied at read time, and we report a \SI{10}{dB} ``data link'' contour throughout. The stored layer can be re-thresholded only to stricter thresholds: cells beyond the ray radius of every candidate hold no value, so a stronger budget than the base case requires a new run (Section~\ref{sec:res_reference}). The downlink threshold $T_{\mathrm{DL}}=N+\mathrm{SNR}_{\mathrm{req}}$ is \SI{-94.4}{dBm}. The downlink received power at a cell at ground distance $d$ from the mast is
\begin{equation}
\Prx(d,f)=P_{\mathrm{BS}}+G_{\mathrm{BS}}+G_{\mathrm{UE}}-L_{\mathrm{fs}}(d,f)-L_{\mathrm{t}}(d,f),
\label{eq:prx}
\end{equation}
where $L_{\mathrm{fs}}=20\log_{10}(4\pi d f/c)$ is the free-space loss and $L_{\mathrm{t}}\ge0$ the terrain excess loss of Section~\ref{sec:tworay}. The uplink is evaluated by reciprocity: the same path with the UE power, so that uplink received power is $\Prx+(P_{\mathrm{UE}}-P_{\mathrm{BS}})$ with $(P_{\mathrm{UE}}-P_{\mathrm{BS}})=\SI{-7}{dB}$, against a base-station threshold $T_{\mathrm{UL}}=\SI{-105.2}{dBm}$ over a \SI{5}{\mega\hertz} cell-edge uplink allocation. Coverage uses the more restrictive direction, i.e.\ a cell is coverable when
\begin{equation}
\Prx\ \ge\ T_{\mathrm{eff}}=\max\bigl\{T_{\mathrm{DL}},\;T_{\mathrm{UL}}-(P_{\mathrm{UE}}-P_{\mathrm{BS}})\bigr\};
\label{eq:threshold}
\end{equation}
with these values the link is downlink-limited and $T_{\mathrm{eff}}=\SI{-94.4}{dBm}$ (a full-carrier uplink assumption would make it uplink-limited at \SI{-92.2}{dBm}). The ray radius is the free-space range to the effective threshold at the lowest carrier $f_{\min}$,
\begin{equation}
r=\frac{c}{4\pi f_{\min}}\,10^{\left(P_{\mathrm{BS}}+G_{\mathrm{BS}}+G_{\mathrm{UE}}-T_{\mathrm{eff}}\right)/20},
\label{eq:radius}
\end{equation}
\SI{22.0}{\kilo\metre} at \SI{1.8}{\giga\hertz}, capped at \SI{100}{\kilo\metre} with the cap reported; the same radius is used for all bands of a run, and the free-space ranges at the other bands are \SI{15.4}{\kilo\metre}, \SI{10.9}{\kilo\metre}, \SI{6.8}{\kilo\metre} and \SI{1.6}{\kilo\metre}. Section~\ref{sec:res_control} shows that this range, not the terrain, is what limits the coverage at every band.

The five carriers are four SFCG-recommended surface
bands~\cite{sfcg32,edwards2023} and the IM-2 demonstration
band~\cite{edwards2023}. They are 1.8\,GHz (3GPP Band~3, flown on
IM-2 but outside the SFCG surface allocations and unlikely to receive
long-term approval because of radio-astronomy concerns), 2.58\,GHz
(n7/38/41, SFCGb1 of~\cite{edwards2023}, expected to carry primary
near-term use), 3.65\,GHz (n48/77/78), 5.855\,GHz (the lower edge of
n47) and 25.375\,GHz (n258).
\subsubsection{Terrain profile and viewshed}
\label{sec:viewshed}
\begin{figure}[!t]
\centering
\includegraphics[width=\columnwidth]{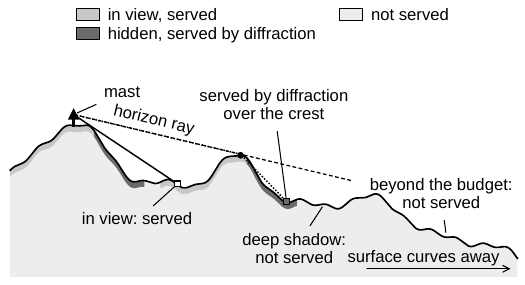}
\caption{Receiver classes are sketched along one ray from a mast on
schematic terrain. Receivers in geometric line of sight are served at
near free-space power. Receivers hidden just behind a crest can still
be served by diffraction over it, at reduced power. Receivers deep in
shadow, or in view but beyond the range of the link budget, are not
served. The engine makes this test on every ray and every cell of the
window.}
\label{fig:concept_profile}
\end{figure}

\begin{figure*}[!t]
\centering
\includegraphics[width=\textwidth]{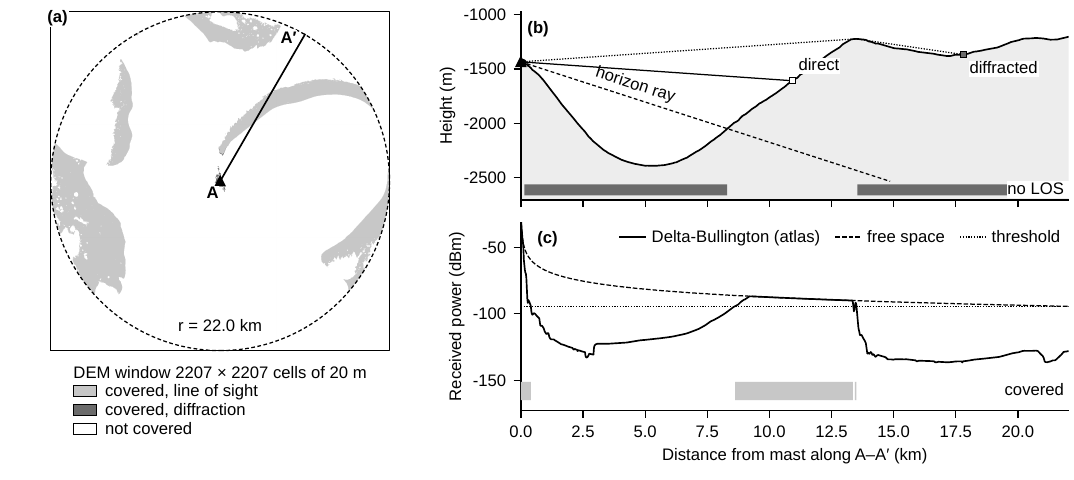}
\caption{The base case of the atlas engine is shown for one candidate
mast (candidate 6430, ridge class, \ang{84.13}S, \ang{131.66}W) at
\SI{1.8}{\giga\hertz} with the atlas link budget (\SI{30}{dBm} base
station, isotropic antennas, \SI{4}{\metre} mast, \SI{1.5}{\metre}
receiver, effective threshold \SI{-94.4}{dBm}, ray radius
\SI{22.0}{\kilo\metre}). Panel (a) gives a plan view of the
$2207\times2207$-cell DEM window (\SI{20}{\metre} cells). Light gray
marks the cells covered in line of sight (\SI{226.8}{\kilo\metre\squared}),
dark gray marks the cells covered by diffraction
(\SI{1.1}{\kilo\metre\squared}), and white marks the cells not covered.
The mast covers \SI{227.9}{\kilo\metre\squared} and sees
\SI{324.3}{\kilo\metre\squared} of the \SI{1526}{\kilo\metre\squared}
disc. Panel (b) shows the terrain along A--A$'$ (azimuth \ang{30},
chosen to cross the visible valley wall) with the sphere drop applied.
It draws the horizon ray through the crest that shadows the valley, a
direct ray to a visible receiver on the far wall, and a diffracted path
over the far crest to a shadowed receiver. The strip marks receivers
without line of sight. Panel (c) shows the received power along
A--A$'$. The curves are the atlas value (Delta-Bullington on every
receiver), free space and the coverage threshold. The strip marks
covered receivers. The far wall is covered in line of sight at
\SIrange{8.6}{13.4}{\kilo\metre}, the valley floor is
\SIrange{25}{35}{dB} below free space, and the only receivers covered
without line of sight lie just beyond the near crest.}
\label{fig:single_mast}
\end{figure*}

For each candidate the engine (Fig.~\ref{fig:level1}, left) reads a square DEM window of half-width one pixel beyond the ray radius (\num{2207} cells wide at \SI{20}{\metre}) and casts $N_{\mathrm{az}}=\num{8192}$ azimuthal rays, a one-pixel arc at the radius, sampled at the pixel spacing. DEM heights $z$ are radial heights above the sphere, so every profile receives the sphere drop before any horizon or diffraction test,
\begin{equation}
z'(d)=z(d)-\frac{d^{2}}{2R},
\label{eq:drop}
\end{equation}
the ``Earth-curvature'' equivalent, here \SI{14}{\metre} at \SI{7}{\kilo\metre} and \SI{140}{\metre} at \SI{22}{\kilo\metre}. With the mast antenna at height $z_{\mathrm{m}}=z'(0)+h_{\mathrm{m}}$, a receiver at distance $d_j$ along the ray is in geometric line of sight (LOS) when its antenna clears the running maximum of the terrain elevation angle,
\begin{equation}
\operatorname{atan2}\bigl(z'(d_j)+h_{\mathrm{r}}-z_{\mathrm{m}},\,d_j\bigr)\ >\ \max_{i<j}\operatorname{atan2}\bigl(z'(d_i)-z_{\mathrm{m}},\,d_i\bigr),
\label{eq:los}
\end{equation}
such that one cumulative-maximum pass along each ray gives the LOS state of every receiver and, as a by-product, the horizon-forming obstruction; the same pass gives the profile needed by the diffraction model. Figure~\ref{fig:concept_profile} sketches the three receiver classes that result along one ray, and Fig.~\ref{fig:single_mast} shows them on real terrain for one candidate. All of this is a batched \texttt{torch} \cite{paszke2019} computation on the GPU. The per-mast cost for all five carriers is \SIrange{14}{29}{\second} on an NVIDIA A30 including the DEM read, and about \SI{6}{\second} per carrier on an RTX~5080 (Section~\ref{sec:computation}).

\subsubsection{Obstructed receivers: Delta-Bullington}
Every receiver, visible or not, is evaluated with the Delta-Bullington method of ITU-R P.1812 Annex~1 \S4.3 \cite{itu1812},
\begin{equation}
L_{\mathrm{t}}=L_{\mathrm{bulla}}+\max\bigl\{L_{\mathrm{sph}}-L_{\mathrm{bulls}},\,0\bigr\},
\label{eq:deltab}
\end{equation}
where $L_{\mathrm{bulla}}$ is Bullington diffraction on the actual profile with the recommendation's empirical term, $L_{\mathrm{bulls}}$ the same on a smooth-Earth profile with least-squares effective antenna heights, and $L_{\mathrm{sph}}$ the first-term spherical-Earth diffraction, all with the lunar radius and the regolith surface admittance. The model is continuous through the radio horizon, which avoids a step at the LOS boundary. Deygout \cite{deygout1966}, Bullington \cite{bullington1947} and single knife-edge (ITU-R P.526 \cite{itu526}) are implemented as alternatives. Deygout edges are frequency-independent and are found once per profile and rescaled by $\sqrt{f}$, so the multi-band run costs little more than a single band. Sensitivity of the coverage class to the model choice is quantified in Section~\ref{sec:validation}.

\subsubsection{Visible receivers: the ground reflection}
\label{sec:tworay}
On visible receivers the deviation from free space is the ground reflection, and Delta-Bullington accounts for it: its spherical-Earth term, evaluated with least-squares effective antenna heights above the fitted profile, is the plane-Earth grazing two-ray loss. On flat ground the Delta-Bullington power reproduces the analytic two-ray field with the Fresnel TM coefficient of the regolith to \SI{0.85}{dB} RMS (\SI{-0.1}{dB} mean, \SIrange{0.2}{2.9}{\kilo\metre}, \SI{1.8}{\giga\hertz}; an engine regression test), and on a \ang{2} plane to within \SI{1}{dB}. A separate two-ray term applied on top of Delta-Bullington on visible receivers, as an earlier version of the engine did, double-counts the reflection at grazing incidence (\SI{10.5}{dB} below the two-ray field on flat ground, \SI{18}{dB} at \SI{2.5}{\kilo\metre}) and is not used. The atlas value is therefore Delta-Bullington on every receiver, visible or not, and the sole terrain-and-ground model of the engine is P.1812 Annex~1 \S4.3 with the lunar radius and the regolith admittance. The parabolic-equation comparison of Section~\ref{sec:val_pe} tests this model on real profiles, including sloped visible paths where the effective-height fit is an approximation.

A second layer, the \emph{no-reflection layer}, is exported alongside free space on visible receivers and Delta-Bullington in shadow. It is an upper bound on the atlas value that shows how much the ground reflection costs on clear paths, and it is the appropriate comparison for ray-tracing results that find no specular path on the decimated terrain (Section~\ref{sec:validation}).

\subsubsection{Products}
At every \SI{20}{\metre} cell and for every carrier $k$ the atlas stores, in both layers, the best-server received power $\Prx$, the maximum over all candidates whose footprint reaches the cell (float32, \SI{-999}{dBm} for cells outside every footprint). The identity of that candidate, whether that candidate is in geometric line of sight of the cell; and the number of candidates whose received power at the cell exceeds the threshold. Two band-independent layers record the number of candidates with line of sight to the cell and the distance to the nearest of them. The coverable mask is $\Prx\ge\SI{-94.4}{dBm}$. As the stored quantity is received power, stricter thresholds and weaker budgets follow by re-thresholding the released layers; a stronger budget or a taller mast changes the ray radius and requires a new run of the engine (Section~\ref{sec:res_reference}). Per candidate, the run records the covered area per band and layer, the LOS area of its disc, the LOS-and-covered area, the LOS area within each band's own free-space range, and the covered fraction of the \SI{2}{\kilo\metre} and \SI{10}{\kilo\metre} EVA discs around it.

\subsection{Computation}
\label{sec:computation}
\subsubsection{Tiling and accumulation}
The \SI{608}{\kilo\metre} disc is divided into pole-centered \SI{40}{\kilo\metre} tiles ($16\times16$ index space, 208 tiles intersecting the \ang{80}S disc), each with a \SI{10}{\kilo\metre} halo used for candidate extraction. A mast's footprint is computed on its own window, independent of tile edges; and merged by the resolver (Fig.~\ref{fig:level1}, right) into per-tile accumulators for every tile it touches, at most nine. Per band the accumulators hold, for the atlas value and for the no-reflection layer alike, the best power, the best identity, the count of masts above threshold and the line-of-sight flag of the best server with $\Prx^{(m)}(c)$ the received power of mast $m$ at cell $c$ that is undefined outside the mast's ray disc and $\ell^{(m)}(c)\in\{0,1\}$ its LOS flag,
\begin{equation}
\begin{aligned}
P^{\star}(c)&=\max_{m}\Prx^{(m)}(c), & m^{\star}(c)&=\arg\max_{m}\Prx^{(m)}(c),\\
n(c)&=\sum_{m}\bigl[\Prx^{(m)}(c)\ge T_{\mathrm{eff}}\bigr], & \ell^{\star}(c)&=\ell^{(m^{\star})}(c);
\end{aligned}
\label{eq:resolver}
\end{equation}
two further layers common to all bands hold the number of masts in line of sight of the cell, $\sum_m\ell^{(m)}(c)$, and the distance to the nearest visible candidate (running minimum). Every operation in (\ref{eq:resolver}) is associative and commutative, so footprints can be accumulated in any order and on any number of workers, and merging shards is exact. No per-candidate footprint is stored (one float32 layer for \num{11827} footprints of five bands on \num{2207}$^2$ windows would be \SI{1.15}{\tera\byte}); a manifest keyed by a hash of the radio parameters, the engine version, the store schema and the candidate table records which candidates are done, and a run is resumable by re-issuing the same command.

\subsubsection{Cluster run}
The whole-pole run was executed on the WPI Turing cluster using Slurm scripts and NVIDIA A30 GPUs as a 16-task job array. Each task processed one shard of \num{739}--\num{740} candidates into its own accumulator directory, with at most two tasks running concurrently. The run reported here is the second pass over the same frozen candidate table, whose SHA-256 is recorded in every manifest together with the engine version and the store schema. The first pass computed the retired two-ray layer alongside the atlas value. The second computes the Delta-Bullington value with the extended store, which holds the count and identity for both layers, the best-server line-of-sight flag, the line-of-sight count and the distance to the nearest visible candidate. Tasks completed in \SI{2.5}{\hour} each, or \SI{12}{\second} per candidate for all five bands including DEM reads, and \SI{40}{GPU\,\hour} in total. Each task flushes its manifest after every candidate and re-queues itself on a stall or at the wall-time signal, so an interrupted task resumes at the next candidate, outlined in the run management of Fig.~\ref{fig:level1}. A CPU-only merge then absorbed the sixteen shard accumulators into a single atlas in \SI{2.7}{\hour}. The merge refuses overlapping candidate sets and differing engine versions, such that counts cannot double and layers cannot mix. It wrote the release bundle of \SI{1.7}{\giga\byte}: pole-wide block-maximum mosaics of every layer at \SI{160}{\metre}, minimum-reduced for the nearest-candidate distance; \SI{20}{\metre} windows around each verified region footprint with an \SI{8}{\kilo\metre} margin; the candidate and coverage tables; and the run parameters and the manifest. The figures of this paper derive from that bundle. The pole-wide statistics derive from the \SI{20}{\metre} accumulators (Section~\ref{sec:statistics}), and the full \SI{20}{\metre} tiles are available from the authors on request.

\subsubsection{Statistics and areas}
\label{sec:statistics}
All pole-wide statistics in this paper are computed exactly on the \SI{20}{\metre} accumulators. The \SI{160}{\metre} release mosaics are block maxima over $8\times8$ cells intended for display. They overstate coverable fractions by \SIrange{2}{8}{} percentage points depending on band, and no statistic in this paper is taken from them. Areas and fractions are given in map (polar stereographic) space unless stated otherwise. The projected disc is \SI{290345}{\kilo\metre\squared} against a spherical cap of \SI{288100}{\kilo\metre\squared}, and the release statistics give both.        %
\section{Results}
\label{sec:results}
This section begins with the validation of the engine against reference solutions and the quantification of its sensitivities (Section~\ref{sec:validation}). The pole-wide results follow (Section~\ref{sec:res_disc}), then the footprints of individual masts (Section~\ref{sec:res_mast}) and the Artemis candidate landing regions (Section~\ref{sec:res_regions}). Those results are then placed in context through three shorter studies, the throughput implied by the stored SNR (Section~\ref{sec:res_throughput}), a power-only control that separates terrain from budget (Section~\ref{sec:res_control}) and a run under the NASA reference budget (Section~\ref{sec:res_reference}). The section closes with a discussion of the scope and limitations of the atlas (Section~\ref{sec:discussion}).
\subsection{Validation and Sensitivity}
\label{sec:validation}

The engine is checked against three independent references a full-wave split-step parabolic-equation solver on real terrain profiles, against an independent ray tracer (NVIDIA Sionna~RT) on line of sight and reflection, and against analytic solutions on canonical geometries. The sensitivity of the coverage \emph{class}, covered or not at the stated threshold, is then quantified with respect to the diffraction model, the DEM product, DEM noise and candidate density, on tiles drawn from every latitude band rather than on a single tile.

\subsubsection{Parabolic-equation reference on real profiles}
\label{sec:val_pe}
A two-dimensional split-step Fourier parabolic-equation (SSPE) solver \cite{levy2000} is used as the reference for both diffraction and ground reflection. The solver is the wide-angle Feit--Fleck propagator \cite{feit1978}, which is exact in vacuum for any range step, with a shift-map treatment of the terrain \cite{barrios1994}. The field is kept in the frame of the current profile segment, where the ground is the flat plane $\zeta=0$, and each change of slope rotates the angular spectrum by the slope change. The ground is an impedance boundary of Leontovich type for vertical polarization,\begin{equation}
\frac{\partial u}{\partial\zeta}+\alpha u=0,\qquad \alpha=\frac{ik\sqrt{\varepsilon_c-1}}{\varepsilon_c},
\label{eq:leontovich}
\end{equation}
and $\varepsilon_c$ from the regolith permittivity and loss tangent of the atlas, implemented with the discrete mixed Fourier transform of Dockery and Kuttler \cite{dockery1996}. The impedance boundary is essential because with $\varepsilon_r=3.7$, the reflection coefficient is \num{0.85} at \ang{2} grazing and \num{0.67} at \ang{5}, angles that sloped ground produces routinely, and a perfectly conducting ground would therefore bias the comparison. The transverse step is $\lambda/4$, which is \SI{4.2}{\centi\metre} at \SI{1.8}{\giga\hertz} and \SI{1.3}{\centi\metre} at \SI{5.855}{\giga\hertz}, which places propagator's Nyquist angle at \ang{90} and no angular truncation is applied in the local frame. The range step is \SI{5}{\metre}. The domain height is the relief of the profile plus \SI{500}{\metre}, with a \SI{300}{\metre} absorbing layer of quadratically rising imaginary index on top, so that energy is never absorbed where later terrain could intercept it. Lunar curvature enters through the earth-flattening index $m-1=z/R$, applied to the raw DEM heights; the sphere drop is thereby counted exactly once. The solver uses no knife-edge approximation and no separate reflection term. Its self-tests, all with the impedance ground unless stated, give an RMS error of \SI{0.13}{dB} for P.526 knife-edge diffraction on a Dirichlet half-plane, \SI{0.14}{dB} for the analytic two-ray field with the Fresnel TM coefficient on flat ground and \SI{0.07}{} and \SI{0.41}{dB} on planes tilted \ang{8.5} and \ang{20}, and \SI{0.10}{dB} for the P.1812 spherical-earth loss over the smooth lunar sphere beyond the \SI{6.0}{\kilo\metre} horizon.

The shift map is exact for reflection on a tilted plane but approximate for diffraction over a convex crest. Against the exact eigenfunction solution for a soft wedge illuminated by a line source, the PE is within \SI{0.7}{dB} RMS for a crest with \ang{2} faces, \SI{0.3}{dB} at \ang{5}, \SI{2.3}{dB} at \ang{7}, \SI{9}{dB} at \ang{10} and fails at \ang{15} (PE too little loss). The slope limit of the treatment is therefore taken as \ang{7} for segment slopes, considerably stricter than the \ang{10}--\ang{15} often quoted, and every receiver of the real-profile comparison is labeled with the steepest segment between it and the mast; statistics are reported for all receivers and for those within the limit. Crests steeper than the limit are covered by the exact wedge solution directly. On the same crests Delta-Bullington differs from it by \SI{+4.1}{dB} (\ang{5} faces), \SI{+1.4}{dB} (\ang{7}), \SI{-4.3}{dB} (\ang{10}) and \SI{-11.0}{dB} (\ang{15}) in the mean, the sign changing as the equivalent knife edge under-predicts the creeping-wave loss of a rounded soft crest; a soft wedge is a bounding case, not regolith, and the numbers are quoted as such.

\begin{figure*}[!t]
\centering
\includegraphics[width=\textwidth]{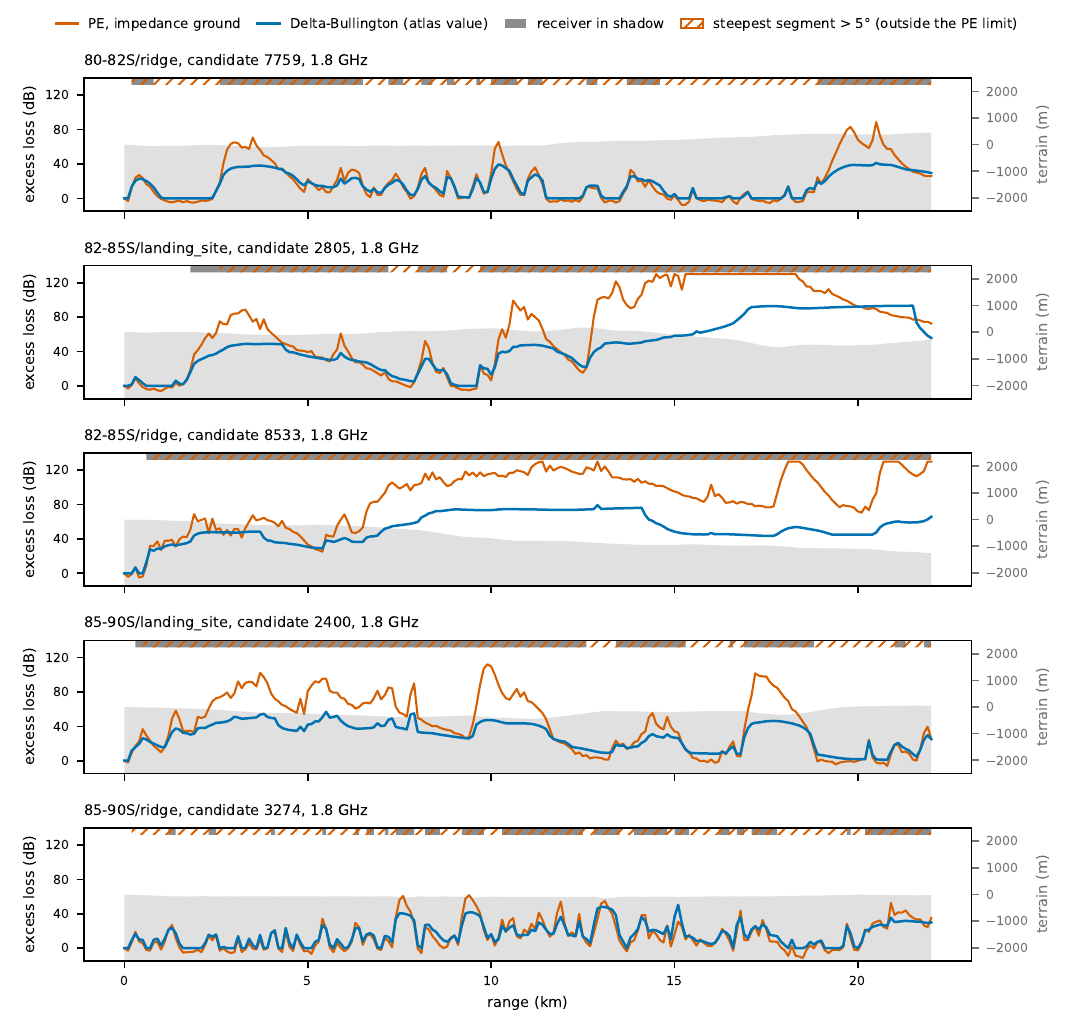}
\caption{Excess loss is shown along real LOLA profiles. Orange marks
the split-step parabolic-equation reference with the impedance ground,
and blue marks Delta-Bullington, the atlas value. The terrain profile
is shaded. The strip along the top marks receivers in shadow (gray) and
receivers behind a segment steeper than \ang{5} (hatched), the limit
adopted for the shadow comparison. One profile is shown per latitude
band and class at \SI{1.8}{\giga\hertz}, in each case at the azimuth
with the most receivers inside the slope limit. The retired two-ray
layer is not drawn because it coincides with the atlas value at this
scale.}
\label{fig:pe_profiles}
\end{figure*}

The comparison runs the atlas engine and the SSPE solver on the same LOLA profiles. Four masts drawn from each latitude band and candidate class that has one (the \ang{80}--\ang{82}S band holds no landing-site candidates: 20 masts), eight azimuths each (160 profiles), at \SI{1.8}{} and \SI{5.855}{\giga\hertz}, with the receiver at \SI{1.5}{\metre}; the PE power is averaged over the receiver height $\pm\SI{0.5}{\metre}$ and over three mast-site heights spanning $\pm\SI{1.3}{\metre}$, so that the comparison is not decided by an unknowable phase at a single geometry. Receivers within \SI{200}{\metre} of the mast and receivers where the PE excess loss exceeds \SI{200}{dB} (below the solver's numerical floor) are excluded, which is why the shadowed counts differ slightly between the two bands. Figure~\ref{fig:pe_profiles} shows representative profiles and Table~\ref{tab:pe} the statistics, for 160 profiles and \num{31369} visible and \num{143146} shadowed receivers, of which \num{2835} and \num{2361} lie within the \ang{7} slope limit.

Two limits are used in reading Table~\ref{tab:pe}. On visible paths the shift map is exact for reflection on a tilted plane. The slope limit, which is a limit on the treatment of \emph{diffraction} over a crest, therefore does not apply there, and the ``visible, all'' rows are the estimate for line-of-sight receivers. In shadow the limit applies, and it is taken at \ang{5} rather than at the \ang{7} of the wedge test. At \ang{7} the PE's own error against the exact wedge, \SI{2.3}{dB}, exceeds the \SI{1.4}{dB} by which Delta-Bullington differs from it there, and a comparison at that limit would judge the model with a reference no better than the model. At \ang{5} the PE is within \SI{0.3}{dB} of the exact solution and Delta-Bullington \SI{4.1}{dB} from it. The \ang{7} rows are kept in the table for reference.

On visible receivers the PE field is \SI{3.9}{dB} \emph{below} Delta-Bullington at \SI{1.8}{\giga\hertz} (RMS \SI{4.6}{dB}, \num{31369} receivers) and \SI{3.3}{dB} below at \SI{5.855}{\giga\hertz} (RMS \SI{4.1}{dB}), with the 5th--95th percentiles of the difference at \SI{-7.9}{} and \SI{+0.1}{dB}. The atlas value is therefore optimistic on clear paths by \SIrange{3}{4}{dB}. Delta-Bullington's effective-height treatment of the grazing reflection, exact on a plane, recovers only part of the destructive interference on the undulating real profiles, where the local grazing angle at the reflection point differs from the plane-fit angle. The retired two-ray layer sits \SI{3.1}{dB} above the PE by the same measure, closer in the mean but for the wrong reason, because its double-counted reflection loss happens to overlap the shortfall of the effective-height fit on these profiles, while on flat ground it is \SI{10}{dB} in error (Section~\ref{sec:tworay}). Delta-Bullington is kept as the atlas value and the bias is propagated rather than ignored. The PE-corrected columns of Table~\ref{tab:disc} re-threshold every cell served in line of sight by the bias measured at its band, \SI{3.9}{dB} at \SI{1.8}{\giga\hertz} and \SI{3.3}{dB} at \SI{5.855}{\giga\hertz}, interpolated between and held constant above, which moves the coverable class of \SI{6.7}{} and \SI{13.1}{\percent} of the visible receivers on the profiles at the two bands and \num{6.7} and \num{16.5} percentage points of the disc (Table~\ref{tab:disc}). In shadow the sign reverses. Behind crests of \ang{5} or less the PE field is \SI{3.6}{dB} \emph{above} Delta-Bullington at \SI{1.8}{\giga\hertz} (RMS \SI{7.9}{dB}, 519 receivers) and \SI{3.5}{dB} above at \SI{5.855}{\giga\hertz} (RMS \SI{9.8}{dB}), the equivalent knife edge over-predicting the loss behind rounded crests, as the wedge test anticipated. The class flips on \SI{7.9}{} and \SI{7.5}{\percent} of those receivers, in the direction of \emph{more} coverage than the atlas shows. Behind steeper crests the wedge test, not the PE, is the reference, and it says the opposite, namely that Delta-Bullington is \emph{optimistic} by \SI{4}{} to \SI{11}{dB} behind \ang{10}--\ang{15} faces. For all shadowed receivers, most of them behind segments steeper than either limit, the PE exceeds Delta-Bullington by \SI{24}{} and \SI{32}{dB} in the mean with RMS values of \SIrange{37}{45}{dB}. Those figures include the receivers on which the shift map itself fails and are quoted only to show why the limit is needed. Three findings follow. First, the atlas is slightly optimistic on the visible paths that carry \SI{97}{\percent} of its coverage, and the PE-corrected fractions of Table~\ref{tab:disc} are the numbers to plan with where that matters. Second, on the few percent that rest on diffraction it is conservative behind gentle crests and optimistic behind steep ones, and Table~\ref{tab:nlos} gives the share of those cells within \SI{5}{} and \SI{10}{dB} of threshold, which is the share at risk. Third, since a \SI{3}{dB} shift moves the free-space range by \SI{30}{\percent}, the in-range covered-to-visible ratio of Table~\ref{tab:cov_los_inrange} is an upper estimate at the edge of the range. All three are statements about the engine, not about the atlas' completeness, and the released line-of-sight flag makes the first a read-time correction, because a planner who wants the PE's answer subtracts the bias from the atlas value on line-of-sight cells.
\begin{table*}[!t]
\caption{Parabolic-equation reference minus the atlas engine on real
LOLA profiles (dB), on visible and shadowed receivers, for all
receivers and for those whose steepest intervening segment is at most
\ang{5} or \ang{7}. A class flip is a receiver whose coverable class
differs between the PE and the atlas value. Bold marks the rows used as
the estimate. All visible receivers qualify because the shift map is
exact for reflection on a tilted plane. Shadowed receivers behind
slopes of at most \ang{5} qualify because there the PE is within
\SI{0.3}{dB} of the exact wedge solution.}
\label{tab:pe}
\centering\scriptsize
\begin{tabular}{@{}llrrrrr@{}}
\toprule
Band & Receivers & $n$ & PE $-$ DB mean & RMS & PE $-$ (DB+two-ray) mean & class flips (\%) \\
\midrule
1.800 GHz & \textbf{visible, all} & 31,369 & \textbf{-3.86} & \textbf{4.61} & -3.13 & \textbf{6.73} \\
1.800 GHz & visible, slopes $\le 5^\circ$ & 1,510 & -3.34 & 4.07 & -2.30 & 0.00 \\
1.800 GHz & \textbf{shadow, slopes $\le 5^\circ$} & 519 & \textbf{+3.57} & \textbf{7.90} & +3.57 & \textbf{7.90} \\
1.800 GHz & shadow, slopes $\le 7^\circ$ & 2,361 & +6.97 & 12.18 & +6.97 & 8.01 \\
1.800 GHz & shadow, all & 143,146 & +24.08 & 36.75 & +24.08 & 1.90 \\
5.855 GHz & \textbf{visible, all} & 31,369 & \textbf{-3.33} & \textbf{4.07} & -2.73 & \textbf{13.08} \\
5.855 GHz & visible, slopes $\le 5^\circ$ & 1,510 & -3.26 & 3.74 & -2.11 & 0.20 \\
5.855 GHz & \textbf{shadow, slopes $\le 5^\circ$} & 519 & \textbf{+3.53} & \textbf{9.79} & +3.53 & \textbf{7.51} \\
5.855 GHz & shadow, slopes $\le 7^\circ$ & 2,361 & +9.08 & 16.12 & +9.08 & 4.11 \\
5.855 GHz & shadow, all & 142,844 & +31.59 & 45.36 & +31.59 & 0.50 \\
\bottomrule
\end{tabular}

\end{table*}

\subsubsection{Ray tracing: line of sight and reflection}
The engine was cross-checked against NVIDIA Sionna~RT 2.1.0 (Mitsuba 3.9.1, CUDA variant) \cite{hoydis2023sionnart,sionnart_docs} on canonical obstacles and on one real tile, T06\_07, which contains the northern half of de~Gerlache crater and the de~Gerlache Rim~2 region. Sionna's scene is a quadric-decimated triangle mesh of the tile (\num{18}~M to \num{2}~M triangles, rims weighted) in a local frame with the sphere drop applied, a regolith \texttt{RadioMaterial}, isotropic antennas and a measurement mesh \SI{1.5}{\metre} above the terrain; paths include line of sight, specular reflection and first-order UTD diffraction. 

On the tile's visible cells Sionna's LOS-only power equals free space to \SI{0.003}{dB}, confirming the calibration of both engines. Sionna finds no specular ground path on the decimated real terrain, where free space minus Sionna-full is \SI{0.002}{dB} on the visible profile receivers. Its full result therefore equals the atlas no-reflection layer on visible cells, and the difference between it and the atlas value is the grazing ground reflection that Delta-Bullington carries and a faceted mesh cannot. On the visible cells of the tile that difference is the excess of the no-reflection layer over the atlas value, which amounts to \SI{11}{} to \SI{14}{\percent} of covered area for a single mast at \SI{1.8}{\giga\hertz} (Section~\ref{sec:res_disc}) and \SIrange{1.6}{2.5}{} percentage points of the pole-wide coverable fraction. The ground reflection itself is validated against the analytic two-ray solution on flat and tilted planes (Section~\ref{sec:tworay}) and against the PE field on real profiles (Section~\ref{sec:val_pe}), not against ray tracing.

\subsubsection{Ray tracing: wedge diffraction on DEM meshes}
On canonical obstacles, a \SI{120}{\metre} crest \SI{3}{\kilo\metre} from a \SI{60}{\metre} mast with 46 receivers \SIrange{0.5}{5}{\kilo\metre} behind it at \SI{1.8}{\giga\hertz}, Sionna reproduces P.526 knife-edge diffraction on a thin screen (\SI{+0.23}{dB} mean, \SI{1.71}{dB} RMS) and on a \ang{60} wedge (\SI{+0.18}{dB}, \SI{1.70}{dB}), but returns no diffracted path for \ang{31} wedges of either orientation or for a Gaussian ridge of \SI{300}{\metre} width. This is the expected behavior of first-order wedge diffraction on a faceted mesh. The diffraction points lie on silhouette edges of the primitive a ray hits, and a smooth DEM crest decimated to facets has no such edge. On the real tile Sionna holds diffracted energy in fewer than \SI{6}{\percent} of shadowed cells and, where it does, sits \SIrange{41}{57}{dB} below Delta-Bullington. Ray tracing is therefore not used as a diffraction reference on lunar terrain, and the PE solver takes that role.

\subsubsection{Sensitivity to the diffraction model}
\label{sec:val_diffraction}
Deygout, Bullington, single knife edge and Delta-Bullington were run on the same masts. On three radial profiles of T06\_07 (a descending slope, a single ridge and a crater interior), Deygout predicts more shadow loss than Delta-Bullington by \SIrange{7}{17}{dB} on the slope and ridge profiles and by \SIrange{30}{38}{dB} deep inside the crater, where Deygout's sub-path sum is known to over-count; the P.526 \S4.2 rounded-obstacle formula exceeds Deygout by \SI{4.9}{dB} on the Gaussian ridge. The coverage class is far less sensitive than the shadow power, shadowed cells are typically \SIrange{10}{100}{dB} below threshold and only \SI{0.8}{\percent} of them lie within $\pm\SI{6}{dB}$ of it. Table~\ref{tab:sensitivity} gives the class-flip and covered-area statistics for 40 masts drawn from the three latitude bands and both classes. Swapping Deygout for Delta-Bullington changes the class of \SIrange{0.55}{0.75}{\percent} of the \SI{22}{\kilo\metre} disc at \SI{1.8}{\giga\hertz} depending on latitude band (95th percentile over masts \SIrange{1.2}{2.0}{\percent}; about \SI{8}{\percent} of the covered union) and \SIrange{0.07}{0.13}{\percent} at \SI{5.855}{\giga\hertz}, and reduces the covered area by \SIrange{6}{8}{\percent} and \SIrange{2}{4}{\percent}. Plain Bullington, which has no spherical-earth term and therefore no grazing reflection, moves the class of \SI{1.4}{\percent} of the disc the other way and adds \SI{22}{\percent} of covered area at \SI{1.8}{\giga\hertz}: the reflection term matters more than the choice of knife-edge construction. The PE comparison above sits between the two, \SIrange{3}{4}{dB} below Delta-Bullington on visible paths and \SI{3.6}{dB} above it in the shadow of gentle crests.

\subsubsection{Sensitivity to the DEM}
\label{sec:val_dem}
Two effects are separated, on masts stratified by the three latitude bands and both classes. The first is the \emph{product difference} between the PDS LDEM\_80S\_20M and the orbit-adjusted PGDA DEM, which is not a random error but a correction of orbit-track residuals. Running the same 40 masts on both products, the LOS state flips for a mean of \SI{0.8}{\percent} of the \SI{22}{\kilo\metre} disc and the coverable class for \SI{0.8}{\percent} at \SI{1.8}{\giga\hertz} (95th percentile \SI{2.3}{\percent}) and \SI{0.2}{\percent} at \SI{5.855}{\giga\hertz}, with a signed covered-area change of \SI{-5}{\percent} in the mean; the change is not uniform. The ridge-class masts of the \ang{82}--\ang{85}S band lose \SI{12}{\percent} of covered area (one of them half) because their site heights differ between the two products, instances of the artifact of Section~\ref{sec:dem_artefacts}, while the landing-site masts, which stand on flat ground, change by \SIrange{3}{4}{\percent}.

The second is the \emph{residual error} of the adjusted product itself. Barker \emph{et al.}\ give the vertical accuracy of the adjusted product as about \SI{1}{\metre} where LOLA tracks are dense and worse in the gaps between tracks \cite{barker2023}. We model it as correlated Gaussian noise with a \SI{200}{\metre} correlation length (the cross-track gap scale) whose amplitude is the PGDA per-pixel height-error map multiplied by \num{1.3}, \num{3} or \num{10} (the first brings the map's median amplitude to the \SI{1.3}{\metre} RMS of the PDS--PGDA product difference, the last is a stress case), added to the DEM including the mast cell, and re-run six masts per latitude band and class (30 masts) with eight realizations each from fixed, recorded seeds. Because the sites were selected as local maxima, zero-mean noise lowers their prominence on average, so the result is reported as a \emph{signed} change in covered area as well as a class-flip fraction. At the level of the published error map ($\times1.3$) the class of \SIrange{0.2}{0.6}{\percent} of the disc changes at \SI{1.8}{\giga\hertz} by latitude band (Table~\ref{tab:sensitivity}; 95th percentile over masts up to \SI{1}{\percent}) and \SIrange{0.05}{0.25}{\percent} at \SI{5.855}{\giga\hertz}, and the covered area of a peak-picked site changes by \SI{-2.5}{\percent} in the outer band, \SI{-2.2}{\percent} in the middle band and \SI{+1.0}{\percent} in the polar band. At three times the map the flips are \SIrange{0.6}{1.7}{\percent} of the disc and the area change \SIrange{-11}{+2.5}{\percent}, and at ten times, \SIrange{2.3}{5.8}{\percent} and \SIrange{-47}{0}{\percent}. The bias is largest in the outer band, where the track gaps are widest and the sites are lowest, and absent in the polar band, where the tracks converge and the error map is small. The pole-wide statement is then that the class of under one percent of cells depends on DEM error at the level of the published error map, with a systematic loss of covered area for peak-picked sites of \SIrange{2}{2.5}{\percent} north of \ang{85}S and a small gain south of it.

\begin{table*}[!t]
\caption{Sensitivity of the coverage class to the model and the DEM, by
latitude band. The entries are the fraction of the
\SI{22}{\kilo\metre} mast disc whose class changes (mean and 95th
percentile over masts, \%) and the signed change of the covered area
(mean, \%) at \SI{1.8}{} and \SI{5.855}{\giga\hertz}. The model and
product rows use 40 masts. The noise rows use 30 masts with eight noise
realizations each. The Bullington row is plain Bullington without the
spherical-Earth term. Bold marks the largest value in each column and,
for the signed area change, the largest magnitude.}
\label{tab:sensitivity}
\centering\scriptsize
\begin{tabular}{@{}llrrrrrr@{}}
\toprule
 & & \multicolumn{3}{c}{1.8 GHz} & \multicolumn{3}{c}{5.855 GHz} \\
\cmidrule(lr){3-5}\cmidrule(lr){6-8}
Perturbation & Band & flip mean & flip p95 & $\Delta$area & flip mean & flip p95 & $\Delta$area \\
\midrule
Deygout vs.\ DB & 80-82S & 0.75 & 1.15 & -7.2 & 0.13 & 0.26 & -3.5 \\
 & 82-85S & 0.66 & 1.40 & -6.2 & 0.08 & 0.19 & -2.2 \\
 & 85-90S & 0.55 & 2.02 & -7.8 & 0.07 & 0.14 & -2.0 \\
Bullington vs.\ DB & 80-82S & 1.74 & 2.60 & +24.7 & 0.27 & 0.50 & +11.0 \\
 & 82-85S & 1.44 & 3.15 & +17.5 & 0.17 & 0.39 & +9.5 \\
 & 85-90S & 1.29 & 4.05 & +25.3 & 0.17 & 0.29 & +13.5 \\
PDS vs.\ PGDA & 80-82S & 0.60 & 1.30 & -1.5 & 0.20 & 0.33 & -1.5 \\
 & 82-85S & 1.13 & 4.71 & -8.1 & 0.28 & 1.23 & -7.0 \\
 & 85-90S & 0.50 & 1.30 & -3.7 & 0.18 & 0.48 & -8.0 \\
DEM noise 1.3x 200m & 80-82S & 0.64 & 0.85 & -2.5 & 0.25 & 0.35 & -4.1 \\
 & 82-85S & 0.43 & 0.95 & -2.2 & 0.09 & 0.20 & -1.9 \\
 & 85-90S & 0.22 & 0.86 & +1.0 & 0.05 & 0.11 & -0.2 \\
DEM noise 3x 200m & 80-82S & 1.67 & 2.18 & -11.1 & 0.64 & 0.80 & -15.8 \\
 & 82-85S & 1.13 & 3.01 & -8.2 & 0.24 & 0.54 & -7.5 \\
 & 85-90S & 0.56 & 2.24 & +2.5 & 0.11 & 0.28 & -1.5 \\
DEM noise 10x 200m & 80-82S & \textbf{5.83} & 9.19 & \textbf{-46.7} & \textbf{1.78} & \textbf{2.60} & \textbf{-53.7} \\
 & 82-85S & 3.48 & 9.13 & -30.7 & 0.76 & 1.74 & -31.8 \\
 & 85-90S & 2.29 & \textbf{9.26} & +0.1 & 0.40 & 0.99 & -9.8 \\
\bottomrule
\end{tabular}

\end{table*}

\subsubsection{Effective resolution of the DEM}
\label{sec:val_effres}
The \SI{20}{\metre} grid of the adjusted product is not its effective resolution everywhere. The LOLA track spacing at \ang{80}S is several hundred meters between orbits, and between tracks the DEM is interpolated. Only close to the pole, where the tracks converge, is every \SI{20}{\metre} cell constrained by measurements. The PGDA release includes per-pixel shot-count, height-error and effective-resolution maps \cite{barker2023}, and the atlas should be read with them, because the LOS state and the diffraction loss of a cell in a track gap are those of the interpolated surface, which is smoother than the true one. Unresolved relief cuts both ways, since an unresolved knoll can hide a receiver or lift a mast, and the coverable fractions at the outer latitudes are therefore likely optimistic rather than bounded. The DEM-noise experiment above is the quantitative form of this caveat. Its amplitude is taken from the height-error map, so it is largest in the track gaps, and its \SI{200}{\metre} correlation length represents the cross-track gap rather than the along-track shot spacing. A second check asks whether the candidate rules themselves respond to the interpolation structure, by comparing the LOLA shot count at candidate cells with that of all cells in the same latitude band, per detector (local maximum and curvature ridge). The interpolation is extensive. \SI{75}{\percent} of all \SI{20}{\metre} cells in the \ang{80}--\ang{82}S band hold no LOLA shot, \SI{67}{\percent} in \ang{82}--\ang{85}S and \SI{43}{\percent} in \ang{85}--\ang{90}S. The candidate cells are measured far more often than the terrain around them. Their zero-shot share is \SI{23}{}, \SI{21}{} and \SI{12}{\percent} in the three bands, \numrange{0.27}{0.31} of the base rate, with \numrange{0.95}{1.5} shots per candidate cell against \numrange{0.30}{0.93} per cell overall. The local-maximum detector, which supplies \SI{95}{\percent} of the ridge class, prefers measured cells the most (ratio \numrange{0.26}{0.29}), and the curvature-ridge detector less so (\numrange{0.52}{0.72}). The preference is the expected one, since a peak that is measured is a peak that exists, and it means that the candidate sites are better constrained than the receivers around them. It also means the candidate table is not a sample of the interpolated surface, which is why the DEM-noise experiment perturbs the mast cell along with the terrain.

\subsubsection{Candidate density}
\label{sec:val_convergence}
The ridge-class candidate set uses a \SI{1}{\kilo\metre} minimum spacing (Section~\ref{sec:candidates}). To test whether the pole-wide coverable fraction has converged with respect to that choice, the extraction and the atlas run were repeated on three tiles, one per latitude band at the band's median candidate density (T01\_08 at \ang{81.4}S, T10\_03 at \ang{83.4}S and T08\_11 at \ang{86.5}S), with a \SI{500}{\metre} spacing and a cap of 2000 sites per tile. A fourth tile, the roughest on which the 150-per-tile cap of the base set binds, was foreseen but does not exist, because no tile reaches the cap (the maximum is 146 on T12\_08, the median 46), and the cap is therefore not a constraint of the released table. Halving the spacing adds only \SIrange{5}{7}{\percent} to the candidate count on the three tiles (62 to 66, 82 to 86, 43 to 46), because the \SI{1}{\kilo\metre} rule already admits nearly every local maximum that passes the prominence, permanent-shadow and slope rules. The spacing rule is not what limits the density; the terrain is. Table~\ref{tab:convergence} gives the result. The coverable fraction of the tile core changes by \SIrange{0.2}{0.8}{} percentage points at \SI{1.8}{\giga\hertz}, by \SIrange{0.3}{2.9}{} at \SI{5.855}{\giga\hertz} and by \SIrange{0.2}{0.5}{} at \SI{25.375}{\giga\hertz}, always upward and with Jaccard indices of the coverable masks of \numrange{0.991}{0.998} at \SI{1.8}{\giga\hertz}. The largest change, \SI{4.6}{} points at \SI{3.65}{\giga\hertz} on the polar tile, is one added site on a rim that overlooks a floor the base set reached only at the low bands. The mean serving count moves by less than \num{0.3}. The pole-wide fractions are therefore converged with respect to the spacing rule to about one percentage point at the low bands and to a few points in the mid bands on rough polar tiles. At \SI{25.375}{\giga\hertz} the fraction is a density statement in the sense that each added site adds its own \SI{1.6}{\kilo\metre} disc.
\begin{table*}[!t]
\caption{Candidate-density convergence. The entries are the change of
the coverable fraction of the tile core when the minimum spacing is
halved from \SI{1}{\kilo\metre} to \SI{500}{\metre}, on one tile per
latitude band at the median candidate density of the band. The Jaccard
column is the overlap of the two coverable masks at
\SI{1.8}{\giga\hertz}. Bold marks the largest change per band, and
italic marks the lowest overlap.}
\label{tab:convergence}
\centering\scriptsize
\begin{tabular}{@{}lrrrrrrrr@{}}
\toprule
Tile & $n$ (1 km) & $n$ (500 m) & \multicolumn{5}{c}{$\Delta$ coverable (pp) at} & Jaccard 1.8 GHz \\
\cmidrule(lr){4-8} & & & 1.8 & 2.58 & 3.65 & 5.855 & 25.375 & \\
\midrule
T01\_08 & 62 & 66 & +0.51 & +0.58 & +0.63 & +0.78 & \textbf{+0.50} & 0.994 \\
T10\_03 & 82 & 86 & +0.16 & +0.18 & +0.24 & +0.28 & +0.19 & 0.998 \\
T08\_11 & 43 & 46 & \textbf{+0.84} & \textbf{+1.93} & \textbf{+4.55} & \textbf{+2.91} & +0.35 & \textit{0.991} \\
\bottomrule
\end{tabular}

\end{table*}

\subsubsection{Internal consistency of the pole-wide run}
Four checks were made on the merged atlas. First, the run reproduces an earlier computation by a previous version of the engine, which stored two layers, Delta-Bullington with the retired two-ray term on top (its atlas value at the time) and Delta-Bullington alone on every receiver, visible or shadowed, which it labeled ``no reflection'' because the separate term was omitted and which is exactly the present atlas value. The comparison is between that second layer and the present one. On the pilot tile T06\_07, on the \num{3.3}~million cells (visible and shadowed) whose best server is the same mast in both runs, the received power agrees to \SI{0.000}{dB} in the mean with no cell above \SI{0.01}{dB}, and the pole-wide coverable, data-link and median-power statistics of Table~\ref{tab:disc} computed from the earlier layer agree with the present ones to the last digit shown. Second, the free-space range at each band for the stated budget (\SI{22.0}{\kilo\metre}, \SI{15.4}{}, \SI{10.9}{}, \SI{6.8}{} and \SI{1.6}{\kilo\metre}) matches the ray radius recorded in the run and the halo widths seen around isolated masts in the mosaics. Third, band-to-band steps of the median best-server power over the region windows (\SI{3.0}{}, \SI{2.6}{}, \SI{3.3}{} and \SI{3.9}{dB} at Amundsen Rim) track the free-space differences of \SI{3.1}{}, \SI{3.0}{}, \SI{4.1}{} and \SI{12.7}{dB}, with the shortfall at the higher bands expected because the best server switches to a nearer mast as distant ones drop below threshold. Fourth, no tile seams or shard-boundary discontinuities are visible in any layer, and the count layer contains no doubled bands, confirming the order-independent merge. These checks establish the consistency of the run. the physics is validated in the preceding subsections.
Throughout, ``coverable'' means best-server received power at or above the effective threshold of \SI{-94.4}{dBm} (SNR $\ge$ \SI{0}{dB}) with every candidate active, and ``data link'' means \SI{10}{dB} above it (\SI{-84.4}{dBm}). Both are properties of the potential-coverage envelope, not of a deployed network. Cells within one pixel of a mast, where the near-field geometry is meaningless, are clipped to \SI{-30}{dBm} before statistics are taken. Every statistic in this section is computed on the \SI{20}{\metre} accumulators of the atlas (Section~\ref{sec:statistics}). The \SI{160}{\metre} mosaics of the release bundle are used only to render the pole-wide maps. Map-projection and sphere-weighted fractions differ by at most \SI{0.1}{} percentage points (last column of Table~\ref{tab:disc}), and all fractions are therefore reported in map space.

\subsection{The whole disc}
\label{sec:res_disc}
\begin{figure*}[t]
\centering
\includegraphics[width=0.93\textwidth,height=0.9\textheight,keepaspectratio]{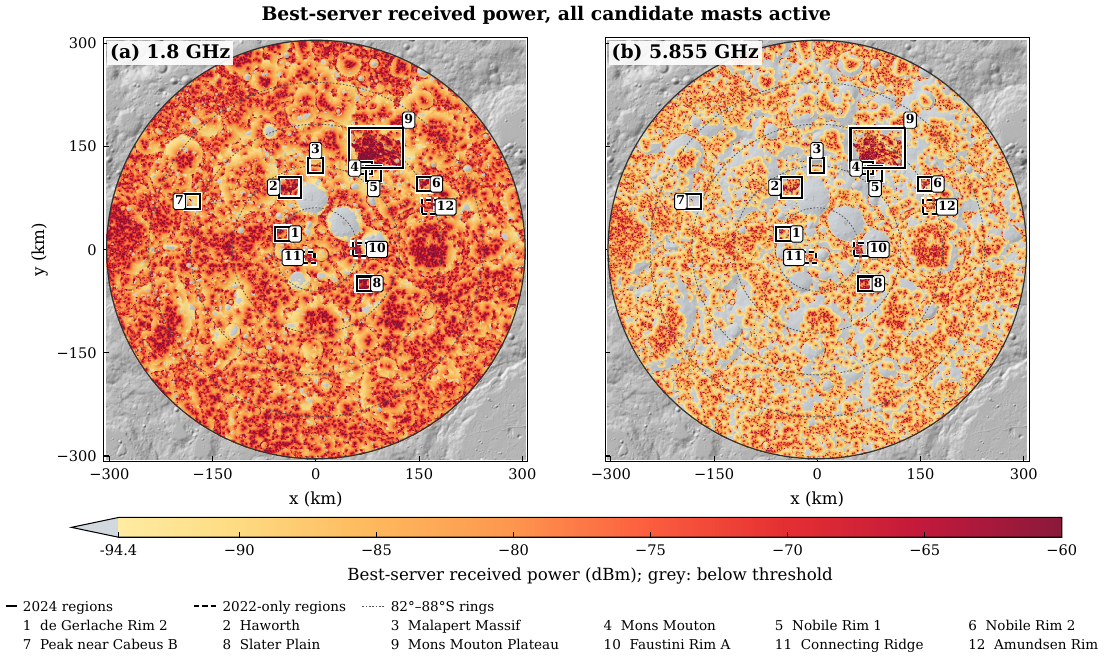}
\caption{Best-server received power is shown over the
\ang{80}S--\ang{90}S disc with every candidate mast active, (a) at
\SI{1.8}{\giga\hertz} and (b) at \SI{5.855}{\giga\hertz}. The maps are
\SI{160}{\metre} block-maximum mosaics rendered at \SI{640}{\metre}.
Gray marks the cells below the \SI{-94.4}{dBm} threshold, which are not
coverable. Black outlines show the nine 2024 Artemis candidate landing
regions as solid lines and the three 2022-only regions with public
footprints as dashed lines. Dotted rings mark \ang{82}S to \ang{88}S.
Footprints are numbered, and the key below the map matches
Table~\ref{tab:regions}. At \SI{1.8}{\giga\hertz} the uncoverable area
is concentrated on, but not confined to, the large permanently shadowed
crater floors near the pole (Table~\ref{tab:psr}). At
\SI{5.855}{\giga\hertz} the plains between ridge clusters also fall
below threshold.}
\label{fig:pole_power}
\end{figure*}

\begin{figure*}[t]
\centering
\includegraphics[width=\textwidth,height=0.9\textheight,keepaspectratio]{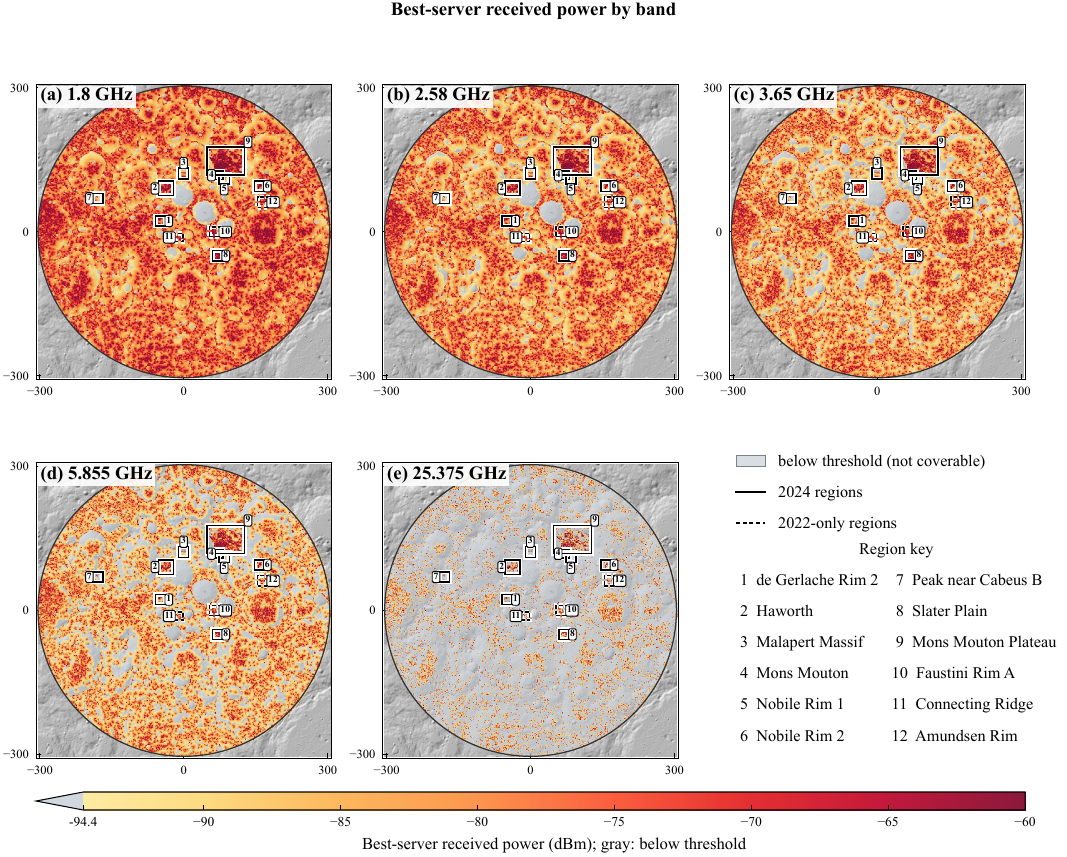}
\caption{Best-server received power for all five bands, same rendering and region numbering as Fig.~\ref{fig:pole_power}, with the symbols and the region key in the lower-right panel. The coverable area shrinks toward the candidate clusters as the free-space range falls from \SI{22.0}{\kilo\metre} at \SI{1.8}{\giga\hertz} to \SI{1.6}{\kilo\metre} at \SI{25.375}{\giga\hertz}.}
\label{fig:pole_allbands}
\end{figure*}

\begin{figure*}[t]
\centering
\includegraphics[width=\textwidth,height=0.9\textheight,keepaspectratio]{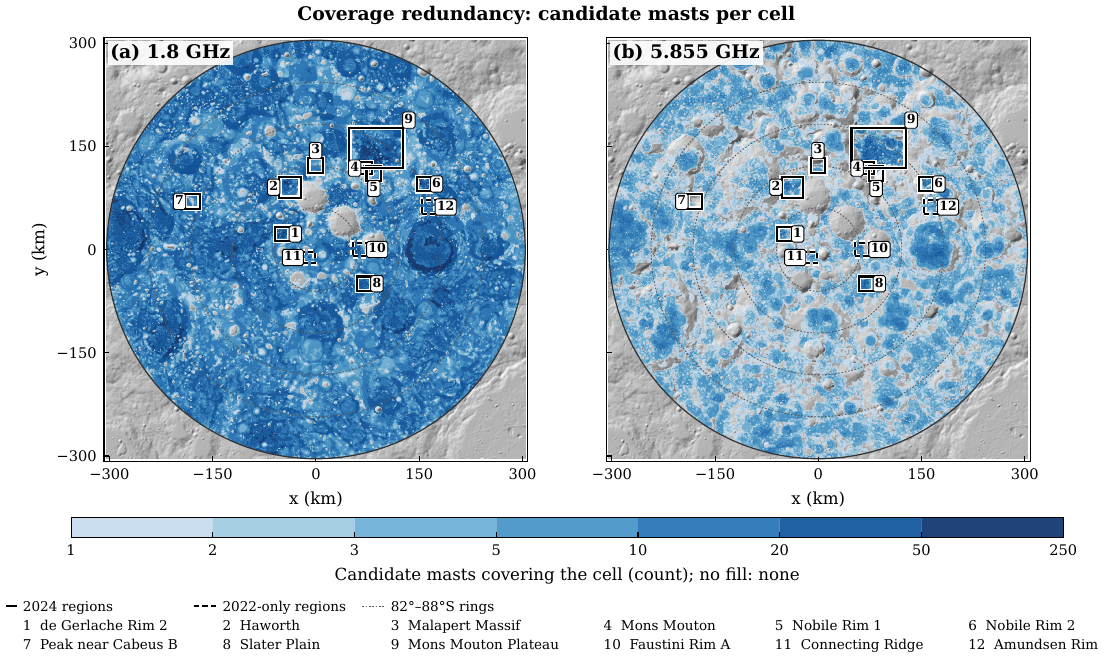}
\caption{Serving redundancy is shown as the number of candidate masts
whose received power at the cell exceeds the threshold, (a) at
\SI{1.8}{\giga\hertz} and (b) at \SI{5.855}{\giga\hertz}. Region
outlines and numbering follow Fig.~\ref{fig:pole_power}. The count
depends on the candidate density (\SI{1}{\kilo\metre} spacing) and is a
measure of siting flexibility, not of a network property. The floors of
illuminated craters such as Amundsen (at about $x=\SI{170}{\kilo\metre}$,
$y=0$) are seen by every rim site, while the plains between ridge
clusters are served by one or two distant masts.}
\label{fig:pole_count}
\end{figure*}

\begin{figure*}[t]
\centering
\includegraphics[width=\textwidth]{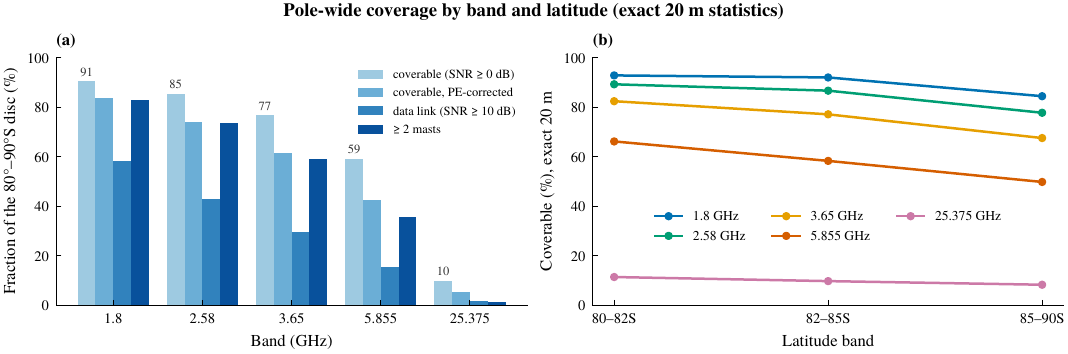}
\caption{Panel (a) gives the coverable, PE-corrected coverable,
data-link and redundant ($\ge2$ masts) fractions of the disc per band,
from the exact \SI{20}{\metre} statistics of
Table~\ref{tab:disc}. Panel (b) gives the coverable fraction by
latitude band (\ang{80}--\ang{82}S, \ang{82}--\ang{85}S,
\ang{85}--\ang{90}S, the scheme used throughout). Coverage falls toward
the pole at every band as the relief and the PSR area grow (exact
values in the text).}
\label{fig:disc_summary}
\end{figure*}

Figure~\ref{fig:pole_power} shows the best-server power at \SI{1.8}{\giga\hertz} and \SI{5.855}{\giga\hertz} over the whole disc, and Fig.~\ref{fig:pole_allbands} shows all five bands. Figure~\ref{fig:pole_count} shows the serving redundancy, Table~\ref{tab:disc} the pole-wide statistics and Fig.~\ref{fig:disc_summary} their dependence on band and latitude. Every cell of the disc lies within the ray radius of at least one candidate, which makes the atlas complete in the sense that the uncoverable area is uncoverable because of terrain and budget.

\begin{table*}[t]
\caption{Pole-wide coverage statistics of the potential-coverage
envelope. All \num{11827} candidates are active, the projected disc is
\SI{290345}{\kilo\metre\squared} (\SI{288100}{\kilo\metre\squared} on
the sphere), and every cell is within the ray radius of at least one
candidate. The statistics are exact at \SI{20}{\metre} on the map
basis, and the last column is sphere-weighted. The PE-corrected entries
are the same two fractions after the cells served in line of sight are
re-thresholded by the bias of the atlas value measured against the
parabolic-equation reference (\SI{3.9}{dB} at \SI{1.8}{\giga\hertz} and
\SI{3.7}{}, \SI{3.5}{}, \SI{3.3}{} and \SI{3.3}{dB} at the higher
bands, Section~\ref{sec:val_pe}). The data-link entries use a threshold
\SI{10}{dB} above the base threshold. The no-reflection entries give
the coverable fraction of the no-reflection layer (free space on
visible cells, Section~\ref{sec:tworay}). The mean-masts entries give
the mean number of candidates above threshold per coverable cell.
Median $\Prx$ and Shannon capacity $C=B\log_2(1+\mathrm{SNR})$ for
$B=\SI{20}{\mega\hertz}$ are over coverable cells.}
\label{tab:disc}
\centering\scriptsize
\begin{tabular}{@{}lrrrrrrrrrr@{}}
\toprule
Band (GHz) & Coverable & Data link & \multicolumn{2}{c}{PE-corrected} & No refl. & $\geq 2$ masts & Mean masts & Median $P_\mathrm{rx}$ & Median $C$ & Coverable, sphere \\
 & (\%) & (\%) & cov. (\%) & data link (\%) & (\%) & (\%) & per cell & (dBm) & (Mbit/s) & (\%) \\
\midrule
1.8 & 90.5 & 58.2 & 83.8 & 38.8 & 92.1 & 82.8 & 10.3 & -81.9 & 85 & 90.5 \\
2.58 & 85.5 & 43.1 & 74.0 & 26.3 & 87.5 & 73.6 & 6.5 & -84.4 & 70 & 85.4 \\
3.65 & 76.7 & 29.4 & 61.7 & 16.9 & 79.2 & 59.1 & 4.2 & -86.4 & 58 & 76.7 \\
5.855 & 59.1 & 15.5 & 42.6 & 8.5 & 61.5 & 35.8 & 2.6 & -88.4 & 47 & 59.0 \\
25.375 & 10.0 & 1.6 & 5.4 & 1.0 & 10.5 & 1.4 & 1.2 & -90.8 & 35 & 10.0 \\
\bottomrule
\end{tabular}

\end{table*}

\paragraph{Reachability is near-universal at the low bands}
At \SI{1.8}{\giga\hertz}, \SI{90.5}{\percent} of the disc is coverable, and \SI{85.5}{\percent} at \SI{2.58}{\giga\hertz} (Table~\ref{tab:disc}). The parabolic-equation reference of Section~\ref{sec:val_pe} finds the engine \SIrange{3}{4}{dB} optimistic on visible paths, and re-thresholding the visible cells by that bias (the ``PE-corrected'' columns of Table~\ref{tab:disc}) gives \SI{83.8}{} and \SI{74.0}{\percent}, with the data-link fraction falling from \SI{58.2}{} to \SI{38.8}{\percent} at \SI{1.8}{\giga\hertz}. The correction bites hardest where the margin is thinnest, since \SI{36}{\percent} of the coverable cells at \SI{1.8}{\giga\hertz} and \SI{74}{\percent} at \SI{5.855}{\giga\hertz} lie within \SI{10}{dB} of threshold. Both sets of numbers are reported throughout. The corrected ones are the conservative planning figures, the atlas values are what the released layers hold, and the line-of-sight flag turns one into the other at read time. The uncoverable remainder, \SI{27549}{\kilo\metre\squared} at \SI{1.8}{\giga\hertz}, is concentrated on, but not confined to, the permanently shadowed terrain near the pole. Table~\ref{tab:psr} gives the split. The LOLA PSR product \cite{barker2023} covers \SI{20829}{\kilo\metre\squared} of the disc (\SI{7.2}{\percent}). Of that area \SI{59.1}{\percent} is coverable at \SI{1.8}{\giga\hertz} and \SI{61.1}{\percent} is in line of sight of at least one candidate, which means that the permanently shadowed floors are, for the most part, \emph{visible} from the permitted sites on their rims and are reached at the low bands. What shadows them from the Sun at grazing incidence does not shadow them from a mast on the rim above. Shackleton is the clearest case. \SI{93.7}{\percent} of its \SI{234}{\kilo\metre\squared} PSR floor is in line of sight of a rim candidate and \SI{92.0}{\percent} is coverable at \SI{1.8}{\giga\hertz}, falling to \SI{38.4}{\percent} at \SI{3.65}{\giga\hertz} and \SI{5.5}{\percent} at \SI{5.855}{\giga\hertz} as the \SIrange{8}{12}{\kilo\metre} rim-to-floor paths drop out of the free-space range. The PSR that is not reached at \SI{1.8}{\giga\hertz}, \SI{41}{\percent} of it, is the part with no line of sight to any candidate, namely the smaller shadowed craters and the deeper parts of the large floors that the \ang{12} slope rule keeps a \SI{4}{\metre} mast away from. That part makes up \SI{31}{\percent} of the uncoverable area. The other \SI{69}{\percent} of the low-band holes lie \emph{outside} permanent shadow, on the shadowed flanks and floors of illuminated craters and on the far sides of massifs. At \SI{1.8}{\giga\hertz}, \SI{81}{\percent} of the holes have no line of sight to any candidate and \SI{19}{\percent} are visible from a candidate but below the budget. Only \SI{0.4}{\percent} of the no-LOS holes lie beyond the ray radius of their nearest candidate, which confirms the completeness of the atlas in the sense of Section~\ref{sec:res_disc}. The statement is conditional on the siting rules. No candidate stands in permanent shadow or on ground steeper than \ang{12}, and the atlas therefore says what a network of small masts on the illuminated uplands reaches. 

\begin{table*}[t]
\caption{Permanently shadowed area and the uncoverable remainder, per
band (exact, \SI{20}{\metre}). The PSR area is taken from the LOLA PSR
product \cite{barker2023} within the disc
(\SI{20829}{\kilo\metre\squared}, \SI{7.2}{\percent}). The PSR LOS
entries count the PSR cells in line of sight of a candidate. The no-LOS
entries count the uncoverable cells with no line of sight to any
candidate, together with the share of those whose nearest candidate
lies beyond the ray radius.}
\label{tab:psr}
\centering\scriptsize
\begin{tabular}{@{}lrrrrrrr@{}}
\toprule
Band & PSR cov. & PSR data link & PSR LOS & Uncoverable & PSR share & no LOS & of which beyond \\
(GHz) & (\% PSR) & (\% PSR) & (\% PSR) & (km$^2$) & (\% of uncov.) & (\% of uncov.) & ray radius (\%) \\
\midrule
1.8 & 59.1 & 24.9 & 61.1 & 27,549 & 31 & 81 & 0 \\
2.58 & 52.5 & 15.7 & 61.1 & 42,213 & 23 & 56 & 0 \\
3.65 & 41.5 & 9.4 & 61.1 & 67,611 & 18 & 36 & 0 \\
5.855 & 25.2 & 4.2 & 61.1 & 118,818 & 13 & 21 & 0 \\
25.375 & 2.4 & 0.2 & 61.1 & 261,403 & 8 & 9 & 0 \\
\bottomrule
\end{tabular}

\end{table*}

\paragraph{How much of the coverage rests on diffraction}
Table~\ref{tab:nlos} separates the two ingredients of the coverage with the best-server line-of-sight flag of the atlas. \SI{91.5}{\percent} of the disc is in line of sight of at least one candidate, and of those cells \SI{98.1}{\percent} are coverable at \SI{1.8}{\giga\hertz}, \SI{64.5}{\percent} at \SI{5.855}{\giga\hertz} and \SI{10.9}{\percent} at \SI{25.375}{\giga\hertz}: the second column is the part of the visible area that the budget reaches, and it is the column that varies with band. The last column is the part of the atlas that rests on the diffraction model: only \SI{2.7}{\percent} of the coverable cells at \SI{1.8}{\giga\hertz} are served by a mast that is not in line of sight, \SI{0.9}{\percent} at \SI{5.855}{\giga\hertz} and \SI{0.3}{\percent} at \SI{25.375}{\giga\hertz}. Diffraction decides the class of a few percent of the disc; the rest is geometry and budget. The last two columns say how much of that diffraction-dependent coverage is at risk: the share of the NLOS-served cells whose margin above threshold is less than \SI{5}{} and \SI{10}{dB}, which is the share that a diffraction model in error by that much would move. The parabolic-equation comparison of Section~\ref{sec:val_pe} finds the engine conservative in shadow behind gentle crests and optimistic by \SIrange{4}{11}{dB} behind crests steeper than \ang{10}, so the at-risk share is the \SI{61}{\percent} of the NLOS-served cells at \SI{1.8}{\giga\hertz} that lie within \SI{10}{dB} of threshold, \SI{1.6}{\percent} of all coverable cells (\SI{0.8}{\percent} at \SI{5.855}{\giga\hertz}); a diffraction model in error by \SI{10}{dB} on those paths could move at most that much of the disc downward.

\begin{table*}[t]
\caption{Line-of-sight content of the coverage, per band (exact,
\SI{20}{\metre}). The first entries give, of the cells in line of sight
of at least one candidate, the fraction above threshold. The
served-without-LOS entries count the coverable cells whose best server
is not in line of sight. The last two columns give the share of those
cells whose margin above threshold is below \SI{5}{} and \SI{10}{dB}.
That share is the part of the diffraction-dependent coverage that a
model error of that size would move.}
\label{tab:nlos}
\centering\scriptsize
\begin{tabular}{@{}lrrrrr@{}}
\toprule
Band (GHz) & LOS to $\geq1$ candidate (\%) & LOS cells coverable (\%) & Served without LOS (\%) & \multicolumn{2}{c}{of which margin $<5$ / $<10$ dB (\%)} \\
\midrule
1.8 & 91.5 & 98.1 & 2.7 & 32 & 61 \\
2.58 & 91.5 & 93.0 & 2.0 & 38 & 67 \\
3.65 & 91.5 & 83.6 & 1.4 & 43 & 73 \\
5.855 & 91.5 & 64.5 & 0.9 & 50 & 81 \\
25.375 & 91.5 & 10.9 & 0.3 & 80 & 98 \\
\bottomrule
\end{tabular}

\end{table*}

\begin{figure}[t]
\centering
\includegraphics[width=\columnwidth]{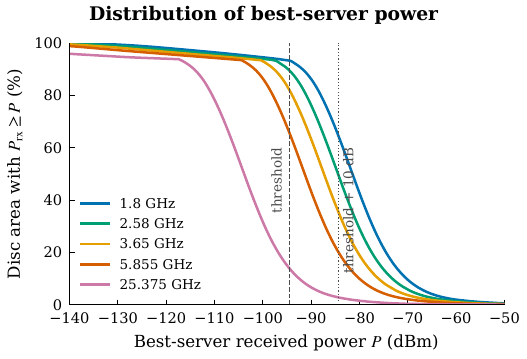}
\caption{Fraction of the disc whose best-server power exceeds $P$, per band. The dashed line is the coverage threshold, the dotted line the \SI{10}{dB} data-link margin. }
\label{fig:power_cdf}
\end{figure}

\paragraph{Coverage falls with the free-space range at every band}
The coverable fraction falls to \SI{76.7}{\percent} at \SI{3.65}{\giga\hertz}, \SI{59.1}{\percent} at \SI{5.855}{\giga\hertz} and \SI{10.0}{\percent} at \SI{25.375}{\giga\hertz} (PE-corrected \SI{61.7}{}, \SI{42.6}{} and \SI{5.4}{\percent}), and the data-link fraction falls faster, from \SI{58.2}{\percent} at \SI{1.8}{\giga\hertz} to \SI{1.6}{\percent} (\SI{38.8}{} to \SI{1.0}{\percent} corrected). The correction moves the \SI{5.855}{\giga\hertz} data-link fraction most, from \SI{15.5}{} to \SI{8.5}{\percent}. In Fig.~\ref{fig:pole_power}(b) the plains between the ridge clusters have turned gray. They were covered at \SI{1.8}{\giga\hertz} by grazing line-of-sight paths from masts \SIrange{10}{20}{\kilo\metre} away, at \SIrange{-92}{-85}{dBm}, and \SI{10}{dB} more free-space loss removes them. The mechanism is the same at every band, including \SI{1.8}{\giga\hertz}. Within its free-space range a mast covers about \SI{90}{\percent} of its visible area whatever the band (Table~\ref{tab:cov_los_inrange}), and the range itself shrinks from \SI{22.0}{\kilo\metre} to \SI{1.6}{\kilo\metre}. The power-only control of Section~\ref{sec:res_control} confirms that the \SI{1.8}{\giga\hertz} coverage is not terrain-limited either, because with \SI{20}{dB} more budget the same masts reach visible terrain tens of kilometers away. The cumulative distribution of Fig.~\ref{fig:power_cdf} makes the same point. The curves are parallel, shifted by the free-space difference, and the threshold line cuts them at very different heights.

\paragraph{Redundancy collapses with frequency, and the millimeter-wave fraction is a density statement}
The mean number of candidate sites able to serve a coverable cell falls from \num{10.3} at \SI{1.8}{\giga\hertz} to \num{2.6} at \SI{5.855}{\giga\hertz} and \num{1.2} at \SI{25.375}{\giga\hertz}, and the fraction of the disc with at least two serving candidates from \SI{82.8}{\percent} to \SI{1.4}{\percent} (Table~\ref{tab:disc}, Fig.~\ref{fig:disc_summary}(a)). At \SI{25.375}{\giga\hertz} essentially every coverable cell has exactly one site that can reach it: the band is a line-of-sight, single-server resource within about \SI{1.5}{\kilo\metre} of a mast, suitable for point-to-point links between mutually visible sites rather than for area coverage. Its pole-wide coverable fraction is therefore a statement about the candidate density (\SI{1}{\kilo\metre} spacing) rather than about the terrain: it grows with the number of candidates rather than converging as the spacing is refined, although slowly: doubling the density adds \numrange{0.2}{0.5} percentage points on the test tiles of Section~\ref{sec:val_convergence}, as most of the added sites stand within \SI{1.6}{\kilo\metre} of an existing one. Figure~\ref{fig:pole_count} shows where the flexibility lies at the low bands: illuminated crater floors ringed by rim candidates (Amundsen, and the large craters of the \ang{80}--\ang{84}S band) are seen by dozens of sites, whereas the plains are single-server territory even at \SI{1.8}{\giga\hertz}.

\paragraph{Latitude}
Coverage falls toward the pole at every band (Fig.~\ref{fig:disc_summary}(b)): at \SI{1.8}{\giga\hertz} from \SI{92.9}{\percent} of the \ang{80}--\ang{82}S band and \SI{92.1}{\percent} of \ang{82}--\ang{85}S to \SI{84.5}{\percent} of the \ang{85}--\ang{90}S band; at \SI{5.855}{\giga\hertz} from \SI{66.2}{} and \SI{58.4}{} to \SI{49.8}{\percent}; at \SI{25.375}{\giga\hertz} from \SI{11.4}{} and \SI{9.7}{} to \SI{8.3}{\percent}. The polar band combines the largest relief, the largest PSR floors (\SI{11.3}{\percent} of its area against \SI{4.8}{\percent} of the outer band) and the fewest illuminated high points: its PSR is \SI{37.8}{\percent} coverable at \SI{1.8}{\giga\hertz} against \SI{73}{\percent} in the two outer bands, and \SI{12.7}{\percent} of its area has no line of sight to any candidate against \SIrange{5.8}{6.3}{\percent}. The outer band holds the densest field of local maxima but the widest gaps between LOLA tracks (Section~\ref{sec:val_effres}).

\paragraph{The ground reflection}
The no-reflection layer (free space on visible cells) exceeds the atlas value by \num{1.6} percentage points of coverable fraction at \SI{1.8}{\giga\hertz}, \num{2.4} at \SI{5.855}{\giga\hertz} and \num{0.5} at \SI{25.375}{\giga\hertz} (Table~\ref{tab:disc}): this is the cost of the grazing ground reflection that the spherical-Earth term of Delta-Bullington carries on clear paths, an upper bound by construction (Section~\ref{sec:tworay}). For a single mast the layer covers \SI{11}{} to \SI{14}{\percent} more area at \SI{1.8}{\giga\hertz} (medians over the landing-site and ridge classes); the pole-wide difference is smaller because the best-server maximum over many masts is taken over the cells where each mast's path is least grazing.

\subsection{Per-mast footprints}
\label{sec:res_mast}
\begin{figure*}[t]
\centering
\includegraphics[width=\textwidth]{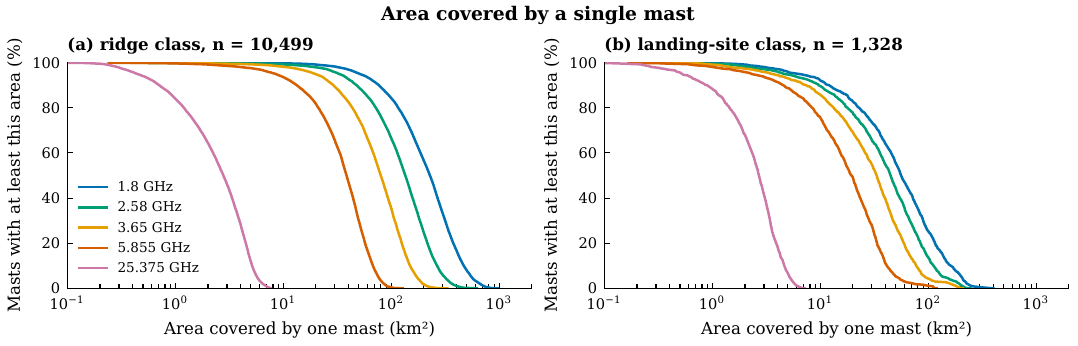}
\caption{Complementary cumulative distribution of the area covered by a single candidate mast, per band, for (a) the ridge class and (b) the landing-site class. The ray disc, $\pi\times(\SI{22.04}{\kilo\metre})^2$, is \SI{1526}{\kilo\metre\squared}.}
\label{fig:mast_area}
\end{figure*}

Figure~\ref{fig:mast_area}, Table~\ref{tab:cand_class} and Table~\ref{tab:cov_los} summarize the coverage of individual sites. A median ridge-class site covers \SI{228}{\kilo\metre\squared} at \SI{1.8}{\giga\hertz} (the best \SI{972}{\kilo\metre\squared}, \SI{64}{\percent} of the \SI{1526}{\kilo\metre\squared} ray disc) and \SI{39}{\kilo\metre\squared} at \SI{5.855}{\giga\hertz}; a median landing-site class site, which by construction stands on flat ground rather than a high point, covers \SI{56}{\kilo\metre\squared} and \SI{19}{\kilo\metre\squared}. The geometric line-of-sight area of a site is small, a median of \SI{18}{\percent} of the ray disc for the ridge class (\SI{269}{\kilo\metre\squared}) and \SI{4}{\percent} (\SI{62}{\kilo\metre\squared}) for the landing-site class. The ratio of covered area to line-of-sight area is informative only when both are measured within the same range: with LOS counted to the \SI{22}{\kilo\metre} ray radius (Table~\ref{tab:cov_los}, retained for comparison) the ratio falls at the higher bands mostly because their free-space range is shorter. Table~\ref{tab:cov_los_inrange} therefore measures LOS and covered area within each band's own free-space range (\SI{22.0}{}, \SI{15.4}{}, \SI{10.9}{}, \SI{6.8}{} and \SI{1.6}{\kilo\metre}), and the result is the same at every band: a ridge-class mast covers a median \SI{89}{\percent} of the terrain it sees within its range (10th--90th percentiles \SIrange{75}{96}{\percent}) and a landing-site mast \SIrange{93}{94}{\percent} (\SIrange{76}{113}{\percent}). The deficit is the grazing visible terrain near the edge of the range, where the ground reflection and the sphere drop take the last few decibels; the excess of a few landing-site masts is diffraction over low ground close to the mast, where the budget has margin to spare. Within the \SI{2}{\kilo\metre} EVA disc, where every path is short, the covered fraction exceeds the visible fraction (\SI{59}{} against \SI{40}{\percent} for the ridge class at \SI{1.8}{\giga\hertz}); over the whole range it does not. The landing-site class keeps a slightly higher ratio because its visible area is compact and close to the mast. For EVA planning the relevant quantity is the covered fraction of the \SI{2}{\kilo\metre} walking disc around a landing-site mast: a median of \SI{49}{\percent} at \SI{1.8}{\giga\hertz} and \SI{35}{\percent} at \SI{5.855}{\giga\hertz} with a \SI{4}{\metre} mast, which is the strongest argument for taller lander-mounted equipment.

\begin{table*}[t]
\caption{Covered area relative to line-of-sight area, both measured
within the free-space range of each band (median and 10th--90th
percentiles over masts). Bold marks the larger class per band.}
\label{tab:cov_los_inrange}
\centering\footnotesize
\begin{tabular}{@{}lrrrrrr@{}}
\toprule
Class & $n$ & \multicolumn{5}{c}{Covered / LOS area, both within the band's free-space range (median; 10th--90th pct.)} \\
\cmidrule(lr){3-7} & & 1.8 GHz (22.0 km) & 2.58 GHz (15.4 km) & 3.65 GHz (10.9 km) & 5.855 GHz (6.8 km) & 25.375 GHz (1.6 km) \\
\midrule
landing-site & 1,328 & \textbf{0.94} (0.76--1.13) & \textbf{0.93} (0.78--1.07) & \textbf{0.93} (0.80--1.03) & \textbf{0.93} (0.83--1.00) & \textbf{0.94} (0.88--0.97) \\
ridge & 10,499 & 0.89 (0.75--0.96) & 0.89 (0.76--0.96) & 0.89 (0.77--0.95) & 0.90 (0.79--0.96) & 0.93 (0.87--0.97) \\
\bottomrule
\end{tabular}

\end{table*}

\begin{table*}[t]
\caption{Per-mast footprint statistics by candidate class. The entries
are the median geometric line-of-sight fraction of the ray disc, the
median covered area, and the median covered fraction of the
\SI{2}{\kilo\metre} EVA disc around the mast. Bold marks the larger
class per column.}
\label{tab:cand_class}
\centering\footnotesize
\begin{tabular}{@{}lrrrrrr@{}}
\toprule
Class & $n$ & LOS frac. & \multicolumn{2}{c}{Covered area per mast (km$^2$)} & \multicolumn{2}{c}{2 km EVA disc covered} \\
\cmidrule(lr){4-5}\cmidrule(lr){6-7} & & (median) & 1.8 GHz median & 5.855 GHz median & 1.8 GHz & 5.855 GHz \\
\midrule
landing-site & 1,328 & 0.04 & 56.0 & 19.5 & 49\% & 35\% \\
ridge & 10,499 & \textbf{0.18} & \textbf{227.9} & \textbf{38.6} & \textbf{59}\% & \textbf{42}\% \\
\bottomrule
\end{tabular}

\end{table*}

\begin{table*}[t]
\caption{Covered area of a single mast relative to its geometric
line-of-sight area within the \SI{22}{\kilo\metre} ray disc, per band
and class (median and 10th--90th percentiles over masts). The LOS area
is counted to \SI{22}{\kilo\metre} at every band, so the fall of the
ratio with frequency mostly reflects the shorter free-space range.
Table~\ref{tab:cov_los_inrange} removes that dependence.}
\label{tab:cov_los}
\centering\footnotesize
\begin{tabular}{@{}lrrrrrrr@{}}
\toprule
Class & $n$ & LOS area & \multicolumn{5}{c}{Covered area / LOS area (median; 10th--90th pct.)} \\
\cmidrule(lr){4-8} & & median (km$^2$) & 1.8 GHz & 2.58 GHz & 3.65 GHz & 5.855 GHz & 25.375 GHz \\
\midrule
landing-site & 1,328 & 62 & 0.94 (0.76--1.13) & 0.80 (0.41--1.04) & 0.61 (0.19--0.99) & 0.34 (0.09--0.91) & 0.04 (0.01--0.23) \\
ridge & 10,499 & 269 & 0.88 (0.74--0.96) & 0.53 (0.31--0.81) & 0.31 (0.14--0.63) & 0.14 (0.05--0.37) & 0.01 (0.00--0.03) \\
\bottomrule
\end{tabular}

\end{table*}

\subsection{The Artemis candidate landing regions}
\label{sec:res_regions}
\begin{figure*}[!p]
\centering
\includegraphics[width=\textwidth,height=0.8\textheight,keepaspectratio]{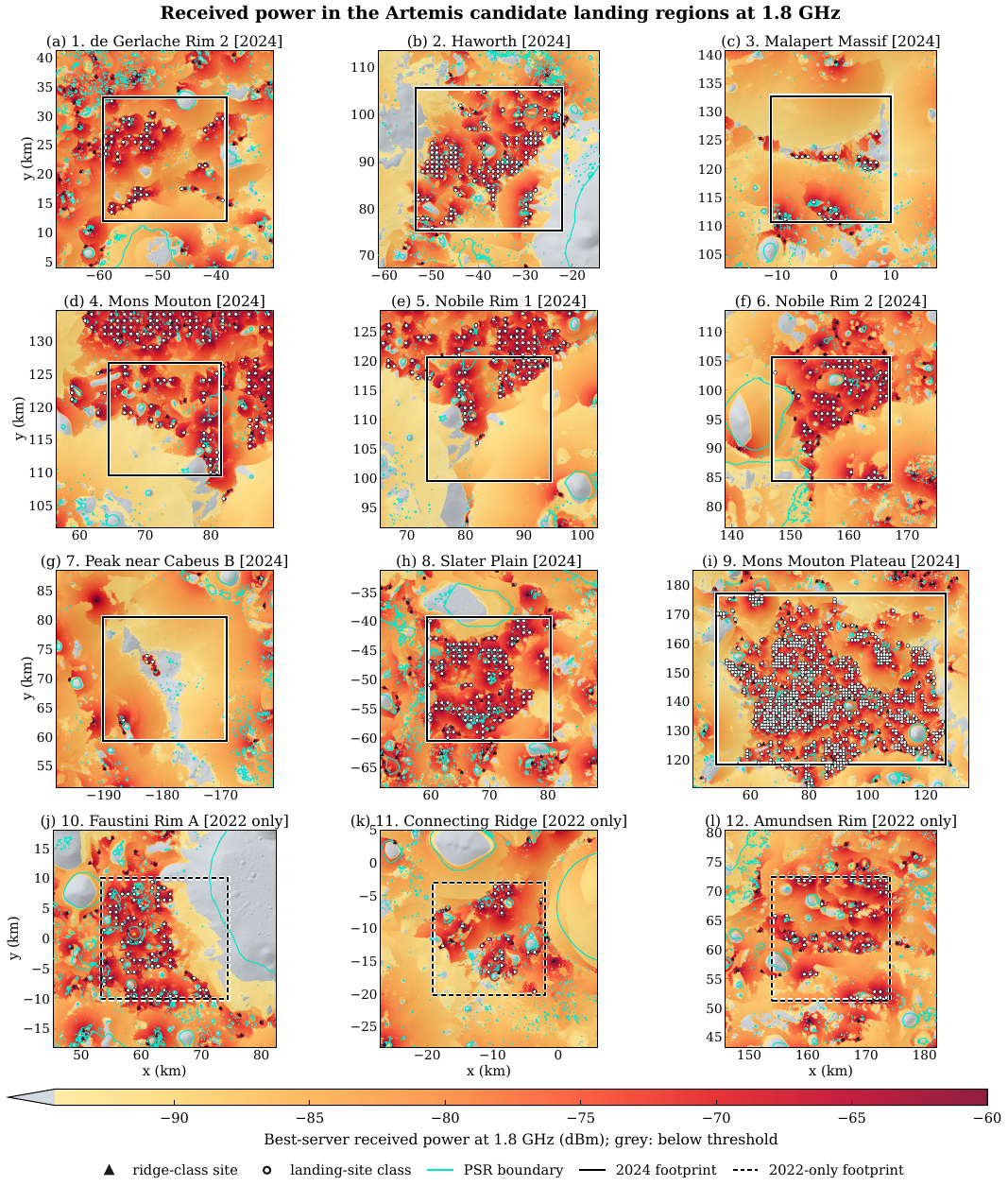}
\caption{Best-server received power at \SI{1.8}{\giga\hertz} is shown
in each Artemis candidate landing region with a public USGS footprint
(\SI{20}{\metre} resolution rendered at \SI{40}{\metre},
\SI{8}{\kilo\metre} margin around the footprint). Panel numbers follow
Fig.~\ref{fig:candidates_pole} and Table~\ref{tab:regions}. Black
triangles mark the ridge-class sites, and white circles mark the
landing-site class. The black box is the footprint, dashed for the
2022-only regions. Panels (a)--(i) are the nine 2024 regions, and
panels (j)--(l) are the 2022-only regions.}
\label{fig:regions}
\end{figure*}

\begin{figure*}[t]
\centering
\includegraphics[width=\textwidth]{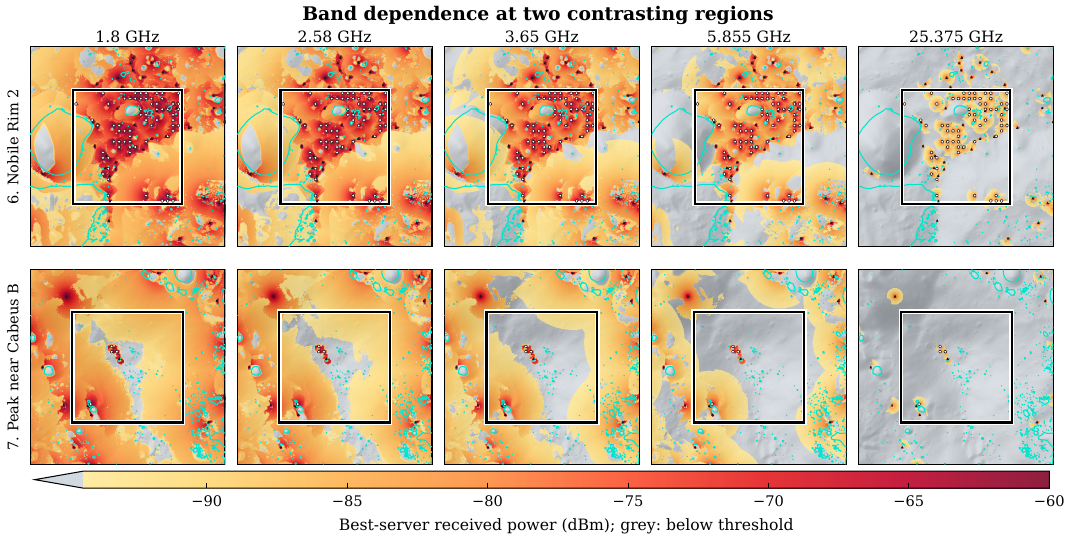}
\caption{Band dependence is shown at two contrasting regions. Nobile
Rim~2 (top, region 6) is \SI{97.7}{\percent} coverable at
\SI{1.8}{\giga\hertz} and \SI{32.0}{\percent} at
\SI{25.375}{\giga\hertz}. Peak near Cabeus~B (bottom, region 7) is
\SI{85.4}{\percent} and \SI{1.9}{\percent} at the same bands. Peak near
Cabeus~B is itself the high ground. Its steep flanks are shadowed from
the sites on the massif and, at the higher bands, beyond the range of
the sites below it (Section~\ref{sec:res_regions}).}
\label{fig:region_bands}
\end{figure*}

\begin{figure*}[t]
\centering
\includegraphics[width=\textwidth]{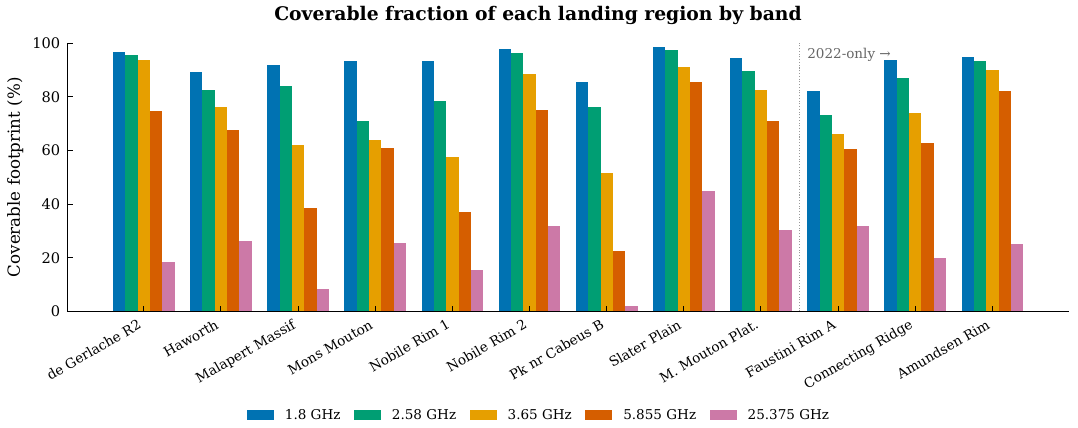}
\caption{The coverable fraction of each region footprint is shown per
band. The nine 2024 regions lie left of the dotted line, and the three
2022-only regions lie right of it.}
\label{fig:regions_chart}
\end{figure*}

Figure~\ref{fig:regions} shows the twelve regions with public footprints at \SI{1.8}{\giga\hertz} at the native \SI{20}{\metre} resolution, Fig.~\ref{fig:region_bands} two contrasting regions across all five bands, Table~\ref{tab:regions} the per-region statistics and Fig.~\ref{fig:regions_chart} their band dependence. Region statistics are exact (\SI{20}{\metre}) and refer to the footprint only, not the margin drawn around it.

\begin{table*}[t]
\caption{Per-region coverage of the footprint (exact, \SI{20}{\metre}).
Numbers follow Fig.~\ref{fig:candidates_pole}. Areas are pixel counts
of the footprint (the two \SI{17}{\kilo\metre} squares give
\SI{292}{\kilo\metre\squared}). The LOS $\le$ \SI{6.8}{\kilo\metre}
entries give the footprint fraction in line of sight of a candidate
within the \SI{5.855}{\giga\hertz} range. The NLOS-served entries count
the coverable cells whose best server is not in line of sight. Bold
marks the highest value in each column and italic the lowest.}
\label{tab:regions}
\centering\footnotesize
\begin{tabular}{@{}rllrrrrrrrrr@{}}
\toprule
No. & Region & List & Area & \multicolumn{5}{c}{Coverable (\%) at} & LOS $\le$ 6.8 km & NLOS-served & Masts/cell \\
\cmidrule(lr){5-9} & & & (km$^2$) & 1.8 & 2.58 & 3.65 & 5.855 & 25.375 & (\% footprint) & 5.855 GHz (\%) & 1.8 GHz \\
\midrule
1 & de Gerlache Rim 2 & 2024 & 446 & 96.5 & 95.4 & \textbf{93.9} & 74.6 & 18.6 & 76.3 & 1.3 & 17.4 \\
2 & Haworth & 2024 & 941 & 89.4 & 82.3 & 76.2 & 67.6 & 26.1 & 68.9 & 2.1 & 11.7 \\
3 & Malapert Massif & 2024 & 468 & 91.8 & 84.0 & 62.0 & 38.5 & 8.2 & 39.5 & 1.5 & \textit{7.2} \\
4 & Mons Mouton & 2024 & 292 & 93.2 & \textit{71.1} & 63.9 & 60.8 & 25.6 & 61.5 & 1.8 & \textbf{31.9} \\
5 & Nobile Rim 1 & 2024 & 446 & 93.3 & 78.4 & 57.4 & 37.0 & 15.5 & 37.4 & 1.9 & 7.6 \\
6 & Nobile Rim 2 & 2024 & 425 & 97.7 & 96.1 & 88.6 & 74.9 & 32.0 & 76.2 & 1.1 & 19.3 \\
7 & Peak near Cabeus B & 2024 & 446 & 85.4 & 76.3 & \textit{51.5} & \textit{22.4} & \textit{1.9} & \textit{22.8} & 1.3 & 11.4 \\
8 & Slater Plain & 2024 & 446 & \textbf{98.5} & \textbf{97.3} & 91.0 & \textbf{85.3} & \textbf{45.0} & \textbf{86.2} & 1.2 & 14.6 \\
9 & Mons Mouton Plateau & 2024 & 4579 & 94.3 & 89.5 & 82.4 & 70.8 & 30.4 & 71.9 & 1.3 & 19.7 \\
10 & Faustini Rim A & 2022 only & 425 & \textit{82.1} & 73.2 & 66.2 & 60.5 & 31.7 & 62.0 & 2.1 & 12.5 \\
11 & Connecting Ridge & 2022 only & 292 & 93.6 & 86.9 & 74.0 & 62.9 & 19.9 & 63.9 & 1.7 & 7.5 \\
12 & Amundsen Rim & 2022 only & 425 & 95.0 & 93.2 & 89.8 & 82.0 & 25.0 & 83.4 & 1.3 & 11.2 \\
\bottomrule
\end{tabular}

\end{table*}

\paragraph{Low-band coverage of the regions is not the problem}
At \SI{1.8}{\giga\hertz} every current region is at least \SI{85.4}{\percent} coverable (Peak near Cabeus~B) and most exceed \SI{93}{\percent}; Slater Plain (\SI{98.5}{\percent}) and Nobile Rim~2 (\SI{97.7}{\percent}) are almost entirely coverable, and \SI{85}{} and \SI{75}{\percent} of their footprints hold a data link. Of the 2022-only regions, Faustini Rim~A is the least coverable of the twelve (\SI{82.1}{\percent}). The uncoverable cells inside the footprints are small shadowed craters and PSR patches.

\paragraph{The regions separate at the higher bands}
At \SI{5.855}{\giga\hertz} the coverable fraction spans \SI{22.4}{\percent} (Peak near Cabeus~B) to \SI{85.3}{\percent} (Slater Plain), and at \SI{25.375}{\giga\hertz} \SI{1.9}{\percent} to \SI{45.0}{\percent}. The ordering is not explained by the number of candidates inside a footprint: Nobile Rim~1 and Mons Mouton have the same landing-site count (53) and the same ridge count (5) yet reach \SI{37.0}{} and \SI{60.8}{\percent} at \SI{5.855}{\giga\hertz}, and Haworth, with the most landing-site candidates after the Plateau (137), sits in the middle at \SI{67.6}{\percent}. What separates the regions at the higher bands is the geometry of the footprint relative to the nearest high ground, and the regions form a continuum from \SI{22}{} to \SI{85}{\percent} rather than two groups. The quantity that orders them is the fraction of the footprint in line of sight of a candidate within the \SI{5.855}{\giga\hertz} free-space range of \SI{6.8}{\kilo\metre} (Table~\ref{tab:regions}, from the nearest-LOS-candidate layer of the atlas): it runs from \SI{22.8}{\percent} at Peak near Cabeus~B to \SI{86.2}{\percent} at Slater Plain, and it equals the \SI{5.855}{\giga\hertz} coverable fraction to within two percentage points in every one of the twelve regions, the difference being the \SIrange{1.1}{2.1}{\percent} of coverable cells served without line of sight. The higher-band coverage of a region is, to this accuracy, its visible fraction within range; the near-identity is a consequence of the small non-line-of-sight share rather than an independent correlation. Where a footprint is a plain or a crater floor overlooked by rim sites within a few kilometers (Slater Plain, Amundsen Rim, de~Gerlache Rim~2, Nobile Rim~2, the Plateau), most of it is in line of sight of several masts at short range and stays above the budget when the range shrinks; where a footprint is itself the high ground, a massif or a rim crest with steep flanks (Malapert Massif, Nobile Rim~1, Peak near Cabeus~B), the sites inside it see one another and the ground below but the flanks and the terrain behind them are shadowed, and the sites outside the footprint that reached the flanks by diffraction at \SI{1.8}{\giga\hertz} are beyond the higher-band range. Mons Mouton fits neither group: it falls from \SI{93.2}{} to \SI{71.1}{\percent} already at \SI{2.58}{\giga\hertz}, the largest step of any region, yet has the highest redundancy (\num{31.9} masts per coverable cell at \SI{1.8}{\giga\hertz}). This is the signature of a footprint of two parts: one overlooked at short range by many sites on the surrounding plateau rim, the other lying below the scarp and reached at \SI{1.8}{\giga\hertz} only by long grazing paths from the rim sites, which the next band already loses; its visible fraction within \SI{6.8}{\kilo\metre} (\SI{61.5}{\percent}) is indeed close to its \SI{5.855}{\giga\hertz} coverage. The data-link fraction at \SI{1.8}{\giga\hertz} anticipates the higher-band ordering: \SI{84.9}{\percent} at Slater Plain against \SI{22.9}{\percent} at Peak near Cabeus~B.

\paragraph{Redundancy differs by a factor of four between regions}
The mean number of serving candidates per coverable cell at \SI{1.8}{\giga\hertz} ranges from \num{7.2} (Malapert Massif) and \num{7.5}--\num{7.6} (Connecting Ridge, Nobile Rim~1) to \num{31.9} (Mons Mouton), the difference being the number of candidate sites on the surrounding high ground that overlook the footprint rather than the number inside it. In the low-redundancy regions the choice of mast site matters and a second mast buys real resilience; in Mons Mouton or the Plateau almost any rim site serves the whole footprint. The count scales with the candidate density (\SI{1}{\kilo\metre} spacing) and is a relative measure between regions. At \SI{25.375}{\giga\hertz} every region has \numrange{1.2}{2.1} serving candidates per cell.

\subsection{Throughput implications}
\label{sec:res_throughput}
The atlas gives SNR, not throughput; the Shannon bound
\begin{equation}
C=B\log_{2}\bigl(1+\mathrm{SNR}\bigr),\qquad \mathrm{SNR}=\Prx-T_{\mathrm{DL}}+\mathrm{SNR}_{\mathrm{req}}\ \text{(dB)},
\label{eq:shannon}
\end{equation}
for the \SI{20}{\mega\hertz} carrier is nevertheless a useful scale (Table~\ref{tab:disc}). The median coverable cell supports \SI{85}{Mbit/s} at \SI{1.8}{\giga\hertz}, \SI{47}{Mbit/s} at \SI{5.855}{\giga\hertz} and \SI{35}{Mbit/s} at \SI{25.375}{\giga\hertz}; practical LTE/NR radios reach \SIrange{60}{75}{\percent} of the bound with a single stream and are modulation-capped above about \SI{20}{dB} SNR, so the practical figures are \SIrange{50}{75}{Mbit/s} near a mast per \SI{20}{\mega\hertz} carrier, scaling with carrier width (the n78 allocation permits \SI{100}{\mega\hertz} carriers) and MIMO order, and \SIrange{5}{10}{Mbit/s} at the coverage edge. The TDD split of the NASA representative scenario \cite{edwards2023} applies to the TDD bands (n41, n48/77/78, n47, n258); Band~3 and n7 are FDD, where uplink and downlink have their own carriers and the uplink rate is set by the \SI{5}{\mega\hertz} cell-edge allocation of the budget rather than by a time split. In either case the uplink, which carries EVA video, has the smaller share.

\subsection{Power-only control: terrain or budget?}
\label{sec:res_control}
Is the coverage at the lowest band terrain-limited, so that more budget would not help, or budget-limited like the higher bands? The control that decides the question is the base link budget plus \SI{20}{dB} at both ends (base station \SI{50}{dBm}, user equipment \SI{43}{dBm}), with the same \SI{4}{\metre} masts, the same \SI{1.5}{\metre} receiver and the same terrain, run with a \SI{100}{\kilo\metre} ray radius on 213 ridge sites, 50 from each of the two outer latitude bands and 113 from the polar band. With \SI{20}{dB} more budget a \SI{4}{\metre} mast serves cells out to the edge of the \SI{100}{\kilo\metre} ray disc: on 72 of the sites the median distance from a covered cell to its serving mast is \SI{37}{\kilo\metre} at \SI{1.8}{\giga\hertz}, \SI{26}{\percent} of the covered cells lie more than \SI{50}{\kilo\metre} from their mast and the 99th percentile is \SI{99}{\kilo\metre}, against a free-space range of \SI{22}{\kilo\metre} in the base case. The terrain that a rim mast sees is not confined to its neighborhood: distant crater walls and massif flanks facing the mast are in line of sight across the intervening basins, and the base budget simply cannot reach them. Table~\ref{tab:power_control} gives the result. With \SI{20}{dB} more budget the median covered area of a mast grows by a factor of \num{5.2} at \SI{1.8}{\giga\hertz}, \num{8.3} at \SI{2.58}{\giga\hertz}, \num{14} at \SI{3.65}{\giga\hertz}, \num{24} at \SI{5.855}{\giga\hertz} and \num{58} at \SI{25.375}{\giga\hertz}, with no dependence on latitude band beyond the noise of 50 sites; the factors fall short of the free-space scaling ($10^{2}$ in area) only because the \SI{100}{\kilo\metre} cap truncates the low bands and because the visible area grows more slowly than the range. The covered-to-visible ratio within the enlarged range is \num{1.11} at \SI{1.8}{\giga\hertz}, \num{1.04} at \SI{2.58}{\giga\hertz} and \numrange{0.96}{0.99} above, against \num{0.89} at every band under the base budget (Table~\ref{tab:cov_los_inrange}): the ratio is a function of the margin at the edge of the range, not of the band, and exceeds one only when the budget has decibels to spare for diffraction over low obstacles near the mast. The conclusion is that no band is terrain-limited. At every band the covered area is the visible area within the free-space range, less about a tenth, and the range is set by the budget. The terrain enters through the visible fraction, which is what the atlas maps.

\begin{table*}[t]
\caption{Power-only control. The entries are the median growth of the
covered area of a mast when the budget is raised by \SI{20}{dB} at both
ends with nothing else changed (\SI{4}{\metre} mast, \SI{1.5}{\metre}
receiver, \SI{100}{\kilo\metre} ray radius), on 213 ridge sites drawn
from the three latitude bands. The values in parentheses give the
median ratio of covered to visible area within the \SI{+20}{dB}
free-space range (capped at \SI{100}{\kilo\metre}). Bold marks the
largest growth per band.}
\label{tab:power_control}
\centering\scriptsize
\begin{tabular}{@{}lrrrrr@{}}
\toprule
Latitude band & \multicolumn{5}{c}{Growth factor of the covered area (covered / visible within range)} \\
\cmidrule(lr){2-6} & 1.8 GHz & 2.58 GHz & 3.65 GHz & 5.855 GHz & 25.375 GHz \\
\midrule
\ang{80}--\ang{82}S ($n=50$) & \textbf{5.7} (1.10) & 8.0 (1.03) & 14.1 (0.99) & \textbf{27.2} (0.96) & 56.2 (0.96) \\
\ang{82}--\ang{85}S ($n=50$) & 4.8 (1.10) & 7.6 (1.03) & 13.0 (0.99) & 23.2 (0.97) & 53.5 (0.96) \\
\ang{85}--\ang{90}S ($n=113$) & 5.2 (1.11) & \textbf{8.4} (1.04) & \textbf{14.2} (0.99) & 25.2 (0.97) & \textbf{61.7} (0.96) \\
all ($n=213$) & \textbf{5.2} (1.11) & \textbf{8.3} (1.04) & \textbf{14.1} (0.99) & \textbf{24.5} (0.97) & \textbf{57.7} (0.96) \\
\bottomrule
\end{tabular}

\end{table*}

\subsection{A reference budget}
\label{sec:res_reference}
The base atlas deliberately uses a small-lander budget and the resulting coverable fractions at the higher bands are low. Because the stored layers hold no value beyond the ray radius of the base run, a stronger budget cannot be obtained by re-thresholding; it requires a second run. The reference run uses the representative NASA scenario \cite{edwards2023} on the landing-site candidates only: base station \SI{50}{dBm} on a \SI{30}{\metre} lander tower, user equipment \SI{23}{dBm} at \SI{3.5}{\metre} (the Lunar Terrain Vehicle), noise figure \SI{2.5}{dB} and a \SI{20}{\mega\hertz} channel in both directions. The scenario is not the ``\SI{20}{dB} stronger budget'' it appears to be. The downlink gains \SI{20}{dB}, but the link is bidirectional and the uplink, with the same \SI{23}{dBm} handset transmitting into a \SI{20}{\mega\hertz} rather than a \SI{5}{\mega\hertz} allocation, loses \SI{2.6}{dB}. In numbers: the scenario is run with its own noise parameters from Table~2 of \cite{edwards2023}, a noise figure of \SI{2.5}{dB} at both ends and a noise temperature of \SI{250}{\kelvin}, which the run takes as the antenna temperature (the base case uses \SI{150}{\kelvin}, Section~\ref{sec:budget}). The base-station noise is then $k\,(T_{\mathrm{ant}}+T_0(F-1))\,B$ with $T_0(F-1)=\SI{226}{\kelvin}$, $T_{\mathrm{sys}}=\SI{476}{\kelvin}$ and $B=\SI{20}{\mega\hertz}$, i.e.\ $N=\SI{-98.8}{dBm}$; the \SI{23}{dBm} uplink therefore closes at a path loss of \SI{121.8}{dB}, against \SI{124.4}{dB} for the downlink-limited base case ($\SI{30}{dBm}-(\SI{-94.4}{dBm})$), and at that path loss the \SI{50}{dBm} downlink delivers \SI{-71.8}{dBm}, the effective threshold stored with the run. With the base case's \SI{150}{\kelvin} instead, the noise would be \SI{-99.8}{dBm}, the threshold \SI{-72.8}{dBm} and the range \SI{18.4}{\kilo\metre}; the conclusion that the scenario is uplink-limited and shorter-ranged than the base case does not depend on the choice. The free-space range at \SI{1.8}{\giga\hertz} is \SI{16.4}{\kilo\metre}, \emph{shorter} than the \SI{22}{\kilo\metre} of the base case. The run was first started with a \SI{100}{\kilo\metre} ray radius, as for the control above; on 134 sites no covered cell lay more than \SI{16.3}{\kilo\metre} from its tower at \SI{1.8}{\giga\hertz} or \SI{5.0}{\kilo\metre} at \SI{5.855}{\giga\hertz}, and the run was completed at the scenario's own range (\SI{16.4}{\kilo\metre} at \SI{1.8}{\giga\hertz} on the effective threshold of \SI{-71.8}{dBm} in downlink terms). What the reference scenario buys is therefore the tower: \SI{30}{\metre} instead of \SI{4}{\metre} raises the visible fraction of the surrounding terrain, and the \SI{3.5}{\metre} vehicle antenna raises it again. Table~\ref{tab:regions_ref} compares the per-region fractions with the base atlas. The two columns are not the same experiment: the base atlas has every candidate active, including the ridge sites on the high ground around each footprint, whereas the reference scenario puts towers on the landable cells inside the footprints only, so the comparison asks what a lander-mounted tower does without the surrounding ridge sites. At \SI{1.8}{\giga\hertz} the reference scenario covers less of every footprint than the base atlas (\SI{83}{} against \SI{97}{\percent} at de~Gerlache Rim~2, \SI{87}{} against \SI{99}{\percent} at Slater Plain), for two reasons that the table separates: the shorter range, and the absence of the ridge sites, which is the whole difference at the three footprints that are themselves the high ground (Malapert Massif \SI{53}{\percent}, Nobile Rim~1 \SI{44}{\percent}, Peak near Cabeus~B \SI{14.5}{\percent}), where the landable cells lie below the crests that shadow them. At \SI{5.855}{\giga\hertz} and \SI{25.375}{\giga\hertz}, however, the towers inside the footprint reach almost the same fraction as the full base atlas (\SI{71}{} against \SI{68}{\percent} at Haworth, \SI{82}{} against \SI{85}{\percent} at Slater Plain, \SI{76}{} against \SI{82}{\percent} at Amundsen Rim; \SI{46}{} against \SI{45}{\percent} at Slater Plain in the millimeter-wave band): at short range the \SI{30}{\metre} tower and the \SI{3.5}{\metre} vehicle antenna raise the visible fraction of the footprint by as much as the base atlas gains from its thousands of surrounding sites, and the data-link fractions at \SI{1.8}{\giga\hertz} are within a few points of the base case in the overlooked regions. The table also shows the margin argument of Section~\ref{sec:res_control} in a second form: in every region and under both budgets the \SI{1.8}{\giga\hertz} data-link column equals the \SI{5.855}{\giga\hertz} coverable column to within a point, because both are the same \SI{10}{dB} step from the \SI{1.8}{\giga\hertz} threshold, one taken in margin and the other in free-space loss. Three base-case conclusions survive the reference scenario unchanged: the \SI{25.375}{\giga\hertz} band remains single-server (\numrange{1.1}{2.3} serving towers per coverable cell); the regions keep their order, Slater Plain, Amundsen Rim and Nobile Rim~2 at the top and Peak near Cabeus~B, Nobile Rim~1 and Malapert Massif at the bottom, because the order is set by the visible fraction (\SI{91}{} to \SI{14.5}{\percent} of the footprint in line of sight of a tower); and the steep regions are poorly served, more so without the ridge sites. What changes is the share of coverage that rests on diffraction, which falls to \SIrange{0.1}{0.5}{\percent} of coverable cells at \SI{5.855}{\giga\hertz} with the towers: a tall tower sees what a small mast has to diffract to.

\begin{table*}[t]
\caption{Per-region coverable fraction of the footprint (exact,
\SI{20}{\metre}). The fractions are given under the base atlas and
under the reference scenario of \cite{edwards2023} on the landing-site
candidates only. The base atlas has every candidate active, with
\SI{4}{\metre} masts, a \SI{30}{dBm} base station and a
\SI{5}{\mega\hertz} uplink. The reference scenario uses a
\SI{30}{\metre} tower, \SI{50}{dBm}, a \SI{3.5}{\metre} user and a
\SI{20}{\mega\hertz} uplink. It is uplink-limited, with an effective
range of \SI{16.4}{\kilo\metre} at \SI{1.8}{\giga\hertz}. The DL 1.8
entries give the data-link fraction at \SI{1.8}{\giga\hertz}. The LOS
entries give the footprint fraction in line of sight of at least one
active candidate. Bold marks the highest value in each column and
italic the lowest.}
\label{tab:regions_ref}
\centering\footnotesize
\begin{tabular}{@{}rlrrrrrrrrrr@{}}
\toprule
No. & Region & \multicolumn{5}{c}{Base atlas, all candidates (\%)} & \multicolumn{5}{c}{Reference scenario, landing-site towers (\%)} \\
\cmidrule(lr){3-7}\cmidrule(lr){8-12} & & 1.8 & DL 1.8 & 5.855 & 25.375 & LOS & 1.8 & DL 1.8 & 5.855 & 25.375 & LOS \\
\midrule
1 & de Gerlache Rim 2 & 96.5 & 75.3 & 74.6 & 18.6 & 96.0 & 82.9 & 57.0 & 57.3 & 20.8 & 84.4 \\
2 & Haworth & 89.4 & 67.1 & 67.6 & 26.1 & 90.4 & 85.7 & 70.7 & 70.9 & 30.3 & 86.4 \\
3 & Malapert Massif & 91.8 & 39.0 & 38.5 & 8.2 & 91.0 & 53.4 & 32.0 & 31.9 & 9.3 & 54.0 \\
4 & Mons Mouton & 93.2 & 60.4 & 60.8 & 25.6 & 92.4 & 81.3 & 62.2 & 62.3 & 27.9 & 82.2 \\
5 & Nobile Rim 1 & 93.3 & 37.4 & 37.0 & 15.5 & 93.6 & 43.9 & 34.1 & 34.1 & 16.4 & 45.1 \\
6 & Nobile Rim 2 & 97.7 & 74.7 & 74.9 & 32.0 & 97.7 & 90.4 & 71.0 & 70.7 & 31.3 & 90.3 \\
7 & Peak near Cabeus B & 85.4 & \textit{22.9} & \textit{22.4} & \textit{1.9} & 86.7 & \textit{14.5} & \textit{8.1} & \textit{7.8} & \textit{1.8} & \textit{14.5} \\
8 & Slater Plain & \textbf{98.5} & \textbf{84.9} & \textbf{85.3} & \textbf{45.0} & \textbf{98.8} & 86.7 & \textbf{81.7} & \textbf{82.0} & \textbf{45.9} & 86.6 \\
9 & Mons Mouton Plateau & 94.3 & 70.8 & 70.8 & 30.4 & 94.1 & 82.9 & 68.7 & 68.6 & 32.4 & 83.4 \\
10 & Faustini Rim A & \textit{82.1} & 59.7 & 60.5 & 31.7 & \textit{86.3} & 78.5 & 61.2 & 61.5 & 34.2 & 80.5 \\
11 & Connecting Ridge & 93.6 & 62.7 & 62.9 & 19.9 & 93.2 & 75.2 & 62.7 & 63.1 & 21.1 & 74.0 \\
12 & Amundsen Rim & 95.0 & 81.8 & 82.0 & 25.0 & 95.0 & \textbf{90.8} & 76.8 & 76.5 & 25.1 & \textbf{91.4} \\
\bottomrule
\end{tabular}

\end{table*}
\subsection{Discussion and Limitations}
\label{sec:discussion}

\paragraph{What the atlas is}
The atlas is an envelope. Every one of \num{11827} candidates transmits at once, and a coverable cell therefore means ``some permitted site could serve this cell with margin'', not ``a deployed network does''. The count layer is the number of such sites and is a measure of siting flexibility that scales with the \SI{1}{\kilo\metre} candidate spacing. Turning the envelope into a design (a minimal set of sites serving a region, a traverse or a set of instruments, with interference between co-channel cells) is a set-cover or reinforcement-learning problem for which the per-candidate table and the best-server identity layer are the inputs. We deliberately stop short of it here.

\paragraph{What is robust and what is not}
The coverable fraction at \SI{1.8}{}--\SI{2.58}{\giga\hertz} is insensitive to the diffraction model (\SI{0.6}{\percent} of the disc, \SI{8}{\percent} of covered cells at the edge), to the no-reflection alternative (\SIrange{1.6}{2.0}{} percentage points) and to the DEM product and DEM noise at the level of the published error map (\SI{0.8}{} and \SI{0.4}{\percent} of cells; Section~\ref{sec:val_dem}), because most shadowed cells sit \SIrange{10}{100}{dB} below threshold and the choices above act on shadowed cells. What the coverable fraction is sensitive to is the level of the visible-path power itself, which none of those choices touches. The \SIrange{3}{4}{dB} by which the atlas value exceeds the parabolic-equation reference on visible paths moves \SI{6.7}{} percentage points of the disc at \SI{1.8}{\giga\hertz} and \SI{16.5}{} at \SI{5.855}{\giga\hertz}, more at the higher bands because a larger share of their coverable cells lies within a few decibels of threshold. That is why the PE-corrected fractions are reported beside the atlas values, and why they, not the no-reflection layer, are the companion to read with the atlas. The reference puts the truth \emph{below} the atlas value on visible paths, and the free-space layer bounds it from the other side. The release includes the line-of-sight flag that makes the correction a read-time operation, and the no-reflection layer with count and identity for each. The \SI{25.375}{\giga\hertz} coverable fraction is in addition a property of candidate density, since its free-space range (\SI{1.6}{\kilo\metre}) is comparable to the candidate spacing. The convergence test of Section~\ref{sec:val_convergence} bounds this dependence to about half a percentage point when the spacing is halved, because the siting rules admit few additional sites.

\paragraph{Validation gap}
The engine is validated against a parabolic-equation solver on real terrain profiles, against ray tracing on line of sight and reflection and against analytic solutions on the flat-ground reflection and the knife edge, but not against measurements on obstructed lunar-like paths, because no such measurements exist for the Moon and the terrestrial analog campaign we are aware of (NASA Desert RATS 2022 \cite{swank2023drats}) has been published only as images. The PE comparison is the strongest available check on the propagation model, and it is a check of one model against another on the same DEM, not against the true surface. Sub-cell roughness, boulders and the interpolated surface between LOLA tracks are absent from both, and the PE's own terrain treatment is trusted only to \ang{5} segment slopes, so on steeper crests the reference is the exact wedge solution rather than the PE. The \SI{5}{\metre} PGDA mosaic of \ang{87}--\ang{90}S \cite{barker2021} allows a further resolution test on the polar regions (Connecting Ridge, de~Gerlache Rim~2, Slater Plain, Haworth) that has not been run.

\paragraph{Assumptions that are not citable}
Four inputs are stated assumptions rather than measurements, namely the antenna temperature (\SI{150}{\kelvin}, \SI{\pm 1}{dB} on the thresholds per \SI{100}{\kelvin}), the receiver noise figures (\SI{7}{} and \SI{3}{dB}, about \SI{1}{dB} on the thresholds per decibel), the antenna mounting heights on the crewed landers (unpublished, the \SI{4}{}, \SI{10}{} and \SI{30}{\metre} cases bracket them), and the density and composition behind the regolith permittivity. The five 2022 regions without a public footprint are excluded rather than reconstructed.

\paragraph{Model omissions}
The engine has no multipath or fading representation, no antenna patterns, no interference, no surface roughness below the DEM cell~\cite{helfenstein1999,bandfield2015}, no lander body blockage, and no time dependence. It is a static, single-user, best-server map. Plasma and dust effects are neglected, as is the \SI{0.8}{\percent} map-scale variation across the disc.

\paragraph{DEM artifacts}
The orbit-track features in the PDS \SI{20}{\metre} product are relevant beyond this work. Any algorithm that selects local maxima on \texttt{LDEM\_80S\_20M} (for communications relays, solar towers or Earth-visibility studies) will prefer them at a rate of about one in ten. The orbit-adjusted product does not select them. The 77 sites that are higher in the adjusted product are inspected individually in the release. All are kept in the run, and 60 are flagged in the released table because they are local maxima in neither product at the \SI{1}{\kilo\metre} scale or stand on cells with no LOLA shot (\SI{36}{\percent} of the 77), which allows a user to exclude them. Where the PDS product must be used, the per-site height difference is an effective filter.

\paragraph{Future work}
The reference budget of Section~\ref{sec:res_reference} shows that a \SI{30}{\metre} lander tower with a vehicle-height user reaches, at \SI{5.855}{\giga\hertz} and above, nearly the same fraction of an overlooked footprint as the full base atlas does with its thousands of surrounding sites, but cannot replace the ridge sites where the footprint is itself the high ground. It also shows that the representative scenario is uplink-limited, leaving its \SI{20}{dB} of extra base-station power unused. The next step is a scenario with the uplink matched to it (a higher-power vehicle radio, or a narrower uplink allocation), a further run of the engine on the same terrain. Coupling the atlas to the LOLA illumination maps gives the solar-power constraint on mast sites, and coupling it to the Earth-visibility maps gives the backhaul constraint. The best-server identity layer supports a set-cover placement study, and the per-region \SI{5}{\metre} runs and a measurement-validated channel model are the next steps toward link-level simulation.

\section{Conclusion and Future Work}
\label{sec:conclusion}

LunarRF is a GPU-based surface-coverage analysis system for terrain-aware
propagation, built from two independent components, a Delta-Bullington
footprint engine that computes the received-power footprint of one
candidate mast at a time and an order-independent resolver that merges each
footprint into the pole-wide best-server, identity, count and line-of-sight
layers. With lunar data, the system has produced the first pole-wide
potential-coverage atlas for 3GPP surface networks at the lunar south pole,
\num{11827} candidate sites on the orbit-adjusted LOLA \SI{20}{\metre}
DEM~\cite{barker2023} at four SFCG-recommended surface bands and the IM-2 band~\cite{sfcg32,edwards2023},
computed in \SI{40}{GPU\,h}. With \SI{4}{\metre} masts and a \SI{30}{dBm}
budget, \SI{90.5}{\percent} of the \ang{80}--\ang{90}S disc is coverable at
\SI{1.8}{\giga\hertz}, \SI{59.1}{\percent} at \SI{5.855}{\giga\hertz} and
\SI{10.0}{\percent} at \SI{25.375}{\giga\hertz}. A ridge-class mast covers
a median \SI{89}{\percent} of its visible terrain within a free-space range
that falls from \SI{22.0}{} to \SI{1.6}{\kilo\metre}, which shows that the budget, not
the terrain, limits coverage at every band. Of the permanently shadowed
area, \SI{59.1}{\percent} is coverable at \SI{1.8}{\giga\hertz} from the
permitted rim sites, and every current Artemis candidate
region~\cite{nasa2024} is at least \SI{85.4}{\percent} coverable at that
band.

Prior lunar propagation studies address one or a few
sites~\cite{edwards2023,shoemakerrim,qin2024}. The atlas gives mission
planners a screening product for every candidate region and an input to
network design, and after landing it identifies from the actual touchdown
point the nearest permitted site with the greatest coverage. The finding
that local-maximum selection on the unadjusted PDS DEM picks orbit-track
artifacts at one ridge site in ten applies to any siting algorithm on that
product, for relays, solar towers or Earth-visibility studies alike.

The atlas is an envelope of what permitted sites could serve, not a
deployed network, and its accuracy is bounded by the engine and the terrain
rather than by the candidate set. Against a parabolic-equation reference on
\num{160} real profiles the engine is \SIrange{3.3}{3.9}{dB} optimistic on
visible paths, a bias the released line-of-sight flag applies at read time. No measurement on obstructed lunar-like paths exists to validate either
model. Between LOLA tracks the DEM is interpolated, and outer-latitude
fractions are therefore likely optimistic. The antenna temperature, lander
mounting heights and regolith density are assumptions rather than
measurements.

Future work includes coupling the atlas to the LOLA illumination and
Earth-visibility maps~\cite{mazarico2011}, set-cover mast placement on the
best-server identity layer, per-region runs on the \SI{5}{\metre} PGDA
mosaic~\cite{barker2021} and validating the propagation model against
measurements on obstructed lunar-analog paths.          %

\section*{Data Availability}
The atlas bundle (\SI{160}{\metre} pole mosaics, \SI{20}{\metre} region windows, candidate and coverage tables, run parameters and manifest), the exact-statistics tables from which every figure and table in this paper is derived, and the validation records (parabolic-equation profiles, sensitivity, convergence and power-control results) are released under CC~BY~4.0 at \url{https://doi.org/10.5281/zenodo.23301640} (about \SI{3}{\giga\byte}). The full \SI{20}{\metre} tiles (\SI{60}{\giga\byte}) are available from the authors on request. The run manifest records the engine version, the configuration hash and the SHA-256 of the candidate table so that the released layers are traceable to one run.

\section*{Acknowledgment}
The whole-pole atlas was computed on the Turing high-performance computing cluster at Worcester Polytechnic Institute, operated by WPI Academic and Research Computing; the authors thank the ARC staff for their support. The Artemis candidate landing region lists follow the NASA releases of 19~August~2022 \cite{nasa2022} and 28~October~2024 \cite{nasa2024} (\url{https://www.nasa.gov/news-release/nasa-provides-update-on-artemis-iii-moon-landing-regions/}); region footprints are the USGS navigational-grid data releases of M.~T.~McClernan \cite{mcclernan2024a,mcclernan2024b}. Elevation, illumination and permanent-shadow products are from the
{LRO} {LOLA} team \cite{mazarico2011, barker2021, barker2023,
neumann2010gdr} via the NASA Planetary Geodynamics Data Archive and the
PDS Geosciences Node. The authors also thank Dr.~Galahad Wernsing and Srivatsan Mukunthraj of
the Wireless Innovation Laboratory, Worcester Polytechnic Institute,
whose careful reading and constructive comments improved an early draft
of this article.

\bibliographystyle{IEEEtran}
\bibliography{refs}

@IEEEtranBSTCTL{IEEEexample:BSTcontrol,
  CTLdash_repeated_names = "no"
}

@inproceedings{edwards2023,
  author    = {Edwards, Bernard L. and Wagner, Raymond and Zemba, Michael and Millard, Wesley and Braham, Stephen and Gifford, Kevin and Somerlock, Oliver},
  title     = {{3GPP} Mobile Telecommunications Technology on the {Moon}},
  booktitle = {Proc. IEEE Aerospace Conference},
  address   = {Big Sky, MT, USA},
  pages     = {1--12},
  year      = {2023},
  doi       = {10.1109/AERO55745.2023.10115746},
  note      = {{NTRS} 20220015268}
}

@article{mazarico2011,
  author  = {Mazarico, E. and Neumann, G. A. and Smith, D. E. and Zuber, M. T. and Torrence, M. H.},
  title   = {Illumination conditions of the lunar polar regions using {LOLA} topography},
  journal = {Icarus},
  volume  = {211},
  number  = {2},
  pages   = {1066--1081},
  year    = {2011},
  doi     = {10.1016/j.icarus.2010.10.030}
}

@article{barker2021,
  author  = {Barker, M. K. and Mazarico, E. and Neumann, G. A. and Smith, D. E. and Zuber, M. T. and Head, J. W.},
  title   = {Improved {LOLA} elevation maps for south pole landing sites: Error estimates and their impact on illumination conditions},
  journal = {Planetary and Space Science},
  volume  = {203},
  pages   = {105119},
  year    = {2021},
  doi     = {10.1016/j.pss.2020.105119}
}

@article{barker2023,
  author  = {Barker, M. K. and Mazarico, E. and Neumann, G. A. and Smith, D. E. and Zuber, M. T. and Head, J. W. and Sun, X.},
  title   = {A New View of the Lunar South Pole from the {Lunar Orbiter Laser Altimeter} ({LOLA})},
  journal = {The Planetary Science Journal},
  volume  = {4},
  number  = {9},
  pages   = {183},
  year    = {2023},
  doi     = {10.3847/PSJ/acf3e1},
  note    = {{Data:} PGDA product 90 (LDEM/LDEC/LPSR\_80S\_20MPP\_ADJ and error maps), doi:10.60903/gsfcpgda-lola-spole}
}

@misc{mcclernan2024a,
  author = {McClernan, M. T.},
  title  = {Down Selected {Artemis III} Candidate Landing Site Navigational Grids},
  howpublished = {U.S. Geological Survey data release},
  year   = {2024},
  doi    = {10.5066/P1MEQ6UK},
  note   = {doi:10.5066/P1MEQ6UK}
}

@misc{mcclernan2024b,
  author = {McClernan, M. T.},
  title  = {Lunar Grid Reference System ({LGRS}) {Artemis III} Candidate Landing Site Navigational Grids in {Artemis} Condensed Coordinate ({ACC}) Format},
  howpublished = {U.S. Geological Survey data release},
  year   = {2024},
  doi    = {10.5066/P13FD9ZD},
  note   = {doi:10.5066/P13FD9ZD}
}

@misc{nasa2022,
  author = {{NASA}},
  title  = {{NASA} Identifies Candidate Regions for Landing Next {Americans} on {Moon}},
  howpublished = {Release 22-089, \url{https://www.nasa.gov/news-release/nasa-identifies-candidate-regions-for-landing-next-americans-on-moon/}},
  year   = {2022},
  month  = aug
}

@misc{nasa2024,
  author = {{NASA}},
  title  = {{NASA} Provides Update on {Artemis III} {Moon} Landing Regions},
  howpublished = {\url{https://www.nasa.gov/news-release/nasa-provides-update-on-artemis-iii-moon-landing-regions/}},
  year   = {2024},
  month  = oct
}

@inproceedings{george2026,
  author    = {George, T. and Rubio, M. A. and Lawrence, S. J. and Curell, P.},
  title     = {Artemis {III} site selection update: Downselection from thirteen to nine candidate regions},
  booktitle = {Proc. 57th Lunar and Planetary Science Conference},
  year      = {2026},
  note      = {Abstract 1901; NTRS 20250011660}
}

@misc{nasa2026,
  author = {{NASA}},
  title  = {{NASA} Adds Mission to {Artemis} Lunar Program, Updates Architecture},
  howpublished = {Release 26-022, \url{https://www.nasa.gov/news-release/nasa-adds-mission-to-artemis-lunar-program-updates-architecture/}},
  year   = {2026},
  month  = feb
}

@misc{im2mission,
  author = {{Nokia}},
  title  = {{Nokia} and {Intuitive Machines} deliver first cellular network to the {Moon}; achieve some key mission objectives},
  howpublished = {Press release, 10 March 2025, \url{https://www.nokia.com/newsroom/nokia-and-intuitive-machines-deliver-first-cellular-network-to-the-moon-achieve-some-key-mission-objectives/}},
  year   = {2025}
}

@article{shoemakerrim,
  author  = {Adebowale, Quadri R. and Ostermann, Shawn},
  title   = {Sub-6-{GHz} {5G} Large-Scale Path Loss Model for {Shoemaker Rim F}: Sensitivity to Transmitter Antenna Pattern},
  journal = {Telecom},
  volume  = {7},
  number  = {1},
  pages   = {21},
  year    = {2026},
  doi     = {10.3390/telecom7010021}
}

@article{pabari,
  author  = {Pabari, Jayesh P. and Acharya, Yashwant B. and Desai, Uday B. and Merchant, Shabbir N. and Krishna, Barla Gopala},
  title   = {Radio frequency modelling for future wireless sensor network on surface of the {Moon}},
  journal = {International Journal of Communications, Network and System Sciences},
  volume  = {3},
  number  = {4},
  pages   = {395--401},
  year    = {2010},
  doi     = {10.4236/ijcns.2010.34050}
}

@article{qin2024,
  author  = {Qin, K. and Gu, S. and Zhang, Z. and Wang, Y. and Liu, Q. and Zhao, C. and Zhang, Q.},
  title   = {Communication tower deployment scheme for lunar surface multi-device communication networks},
  journal = {Systems Engineering and Electronics},
  volume  = {46},
  number  = {2},
  pages   = {729--739},
  year    = {2024},
  doi     = {10.12305/j.issn.1001-506X.2024.02.37},
  note    = {{In} Chinese}
}

@inproceedings{lordos2020,
  author    = {Lordos, George and others},
  title     = {Autonomously Deployable Tower Infrastructure for Exploration and Communication in Lunar Permanently Shadowed Regions},
  booktitle = {Proc. AIAA ASCEND},
  year      = {2020},
  note      = {{AIAA} 2020-4109},
  doi       = {10.2514/6.2020-4109}
}

@misc{im2install,
  author = {{Nokia}},
  title  = {{Nokia}'s cellular network ready for {Moon} as {Intuitive Machines} completes final lunar lander installation},
  howpublished = {Press release, 7 January 2025, \url{https://www.nokia.com/newsroom/nokias-cellular-network-ready-for-moon-as-intuitive-machines-completes-final-lunar-lander-installation/}},
  year   = {2025},
  note   = {{The} base station is mounted on an upper panel of the Nova-C lander}
}

@article{barrios1994,
  author    = {Barrios, Amalia E.},
  title     = {A terrain parabolic equation model for propagation in the troposphere},
  journal   = {IEEE Transactions on Antennas and Propagation},
  volume    = {42},
  number    = {1},
  pages     = {90--98},
  year      = {1994},
  doi       = {10.1109/8.272306}
}

@article{dockery1996,
  author  = {Dockery, G. Daniel and Kuttler, James R.},
  title   = {An improved impedance-boundary algorithm for {Fourier} split-step solutions of the parabolic wave equation},
  journal = {IEEE Transactions on Antennas and Propagation},
  volume  = {44},
  number  = {12},
  pages   = {1592--1599},
  year    = {1996},
  doi     = {10.1109/8.546245}
}

@misc{hoydis2023sionnart,
  author = {Hoydis, Jakob and {Ait Aoudia}, Fay{\c{c}}al and Cammerer, Sebastian and Nimier-David, Merlin and Binder, Nikolaus and Marcus, Guillermo and Keller, Alexander},
  title  = {Sionna {RT}: Differentiable Ray Tracing for Radio Propagation Modeling},
  howpublished = {arXiv:2303.11103},
  year   = {2023}
}

@misc{sionnart_docs,
  author = {{NVIDIA}},
  title  = {Sionna {RT} 1.x/2.x documentation: Radio propagation, Diffraction},
  howpublished = {\url{https://nvlabs.github.io/sionna/rt/}},
  year   = {2025},
  note   = {{Version} 2.1.0 used; first-order UTD wedge diffraction on scene edges}
}

@book{levy2000,
  author    = {Levy, Mireille},
  title     = {Parabolic Equation Methods for Electromagnetic Wave Propagation},
  series    = {IEE Electromagnetic Waves Series},
  volume    = {45},
  publisher = {Institution of Electrical Engineers},
  address   = {London},
  year      = {2000}
}

@article{penaasensio2025,
  author  = {Pe{\~n}a-Asensio, Eloy and Neira-Acosta, Miguel and S{\'a}nchez-Lozano, Juan Miguel},
  title   = {Evaluating potential landing sites for the {Artemis III} mission using a multi-criteria decision making approach},
  journal = {Acta Astronautica},
  volume  = {226},
  pages   = {469--478},
  year    = {2025},
  doi     = {10.1016/j.actaastro.2024.10.049},
  note    = {{Part} 1; arXiv:2406.19863}
}

@article{olhoeft1975,
  author  = {Olhoeft, G. R. and Strangway, D. W.},
  title   = {Dielectric properties of the first 100 meters of the {Moon}},
  journal = {Earth and Planetary Science Letters},
  volume  = {24},
  number  = {3},
  pages   = {394--404},
  year    = {1975},
  doi     = {10.1016/0012-821X(75)90146-6}
}

@article{helfenstein1999,
  author  = {Helfenstein, P. and Shepard, M. K.},
  title   = {Submillimeter-scale topography of the lunar regolith},
  journal = {Icarus},
  volume  = {141},
  number  = {1},
  pages   = {107--131},
  year    = {1999},
  doi     = {10.1006/icar.1999.6160}
}

@article{bandfield2015,
  author  = {Bandfield, J. L. and Hayne, P. O. and Williams, J.-P. and Greenhagen, B. T. and Paige, D. A.},
  title   = {Lunar surface roughness derived from {LRO} {Diviner} {Radiometer} observations},
  journal = {Icarus},
  volume  = {248},
  pages   = {357--372},
  year    = {2015},
  doi     = {10.1016/j.icarus.2014.11.009}
}

@techreport{morabito2006,
  author      = {Morabito, D. D.},
  title       = {Lunar Noise-Temperature Increase Measurements at {S}-Band, {X}-Band, and {Ka}-Band Using a 34-Meter-Diameter Beam-Waveguide Antenna},
  institution = {Jet Propulsion Laboratory},
  number      = {IPN Progress Report 42-166},
  address     = {Pasadena, CA, USA},
  year        = {2006},
  month       = aug,
  note        = {\url{https://ipnpr.jpl.nasa.gov/progress_report/42-166/166C.pdf}}
}

@article{wei2019,
  author  = {Wei, G. and others},
  title   = {A New Method for Simulation of Lunar Microwave Brightness Temperatures and Evaluation of {Chang'E-2} {MRM} Data Using Thermal Constraints From {Diviner}},
  journal = {Journal of Geophysical Research: Planets},
  volume  = {124},
  number  = {6},
  pages   = {1433--1450},
  year    = {2019},
  doi     = {10.1029/2018JE005858}
}

@misc{itu1812,
  author = {{ITU-R}},
  title  = {Recommendation {ITU-R P.1812}: A path-specific propagation prediction method for point-to-area terrestrial services in the frequency range 30 {MHz} to 6000 {MHz}},
  howpublished = {International Telecommunication Union, Geneva},
  note   = {{Rec.} P.1812-7 (2023), Annex 1, \S4.3},
  year   = {2023}
}

@misc{itu526,
  author = {{ITU-R}},
  title  = {Recommendation {ITU-R P.526}: Propagation by diffraction},
  howpublished = {International Telecommunication Union, Geneva},
  note   = {{Rec.} P.526-15 (2019)},
  year   = {2019}
}

@misc{hoydis2022sionna,
  author = {Hoydis, Jakob and Cammerer, Sebastian and {Ait Aoudia}, Fay{\c{c}}al and Vem, Avinash and Binder, Nikolaus and Marcus, Guillermo and Keller, Alexander},
  title  = {Sionna: An Open-Source Library for Next-Generation Physical Layer Research},
  howpublished = {arXiv:2203.11854},
  year   = {2022}
}

@article{wang2025bictr,
  author  = {Wang, H. Y. and Wyglinski, A. M.},
  title   = {Barren, irregular, chaotic terrain ring model for lunar {5G} applications},
  journal = {IEEE Open J. Commun. Soc.},
  volume  = {6},
  pages   = {9523--9533},
  year    = {2025},
  doi     = {10.1109/OJCOMS.2025.3630145}
}

@inproceedings{chizhik2020,
  author    = {Chizhik, Dmitry and Moilanen, Jani and Klein, Siegfried and Maestro, Luis and Valenzuela, Reinaldo A.},
  title     = {Analytic propagation approximation over variable terrain and comparison to data},
  booktitle = {Proc. 14th European Conference on Antennas and Propagation (EuCAP)},
  address   = {Copenhagen, Denmark},
  year      = {2020},
  doi       = {10.23919/EuCAP48036.2020.9136055}
}

@techreport{foore2007,
  author      = {Foore, Larry and Ida, Nathan},
  title       = {Path loss prediction over the lunar surface utilizing a modified {Longley-Rice} irregular terrain model},
  institution = {NASA},
  number      = {NASA/TM-2007-214825 (E-16017)},
  address     = {Cleveland, OH, USA},
  year        = {2007},
  note        = {{Presented} at the Wireless for Space Workshop, Colorado Springs, CO; NTRS 20070025224}
}

@inproceedings{hwu2008,
  author    = {Hwu, Shian and Upanavage, Matthew and Sham, Catherine},
  title     = {Lunar surface propagation modeling and effects on communications},
  booktitle = {Proc. 26th AIAA International Communications Satellite Systems Conference (ICSSC)},
  address   = {San Diego, CA, USA},
  year      = {2008},
  note      = {{NTRS} 20080015374}
}

@techreport{swank2023drats,
  author      = {Swank, Aaron J. and Zemba, Michael J. and Pember, Taylor C. and Seidl, Cameron M. and Johnson, Kerry J. and Varaljay, Nicholas C. and Yingling, Aaron J. and Caldwell, Sean W.},
  title       = {Lunar {LTE} Studies {DRATS} 2022 Report},
  institution = {NASA Glenn Research Center},
  address     = {Cleveland, OH, USA},
  number      = {NASA/TM-20230014072},
  year        = {2023},
  month       = oct
}

@article{feit1978,
  author  = {Feit, M. D. and Fleck, J. A.},
  title   = {Light propagation in graded-index optical fibers},
  journal = {Applied Optics},
  volume  = {17},
  number  = {24},
  pages   = {3990--3998},
  year    = {1978},
  doi     = {10.1364/AO.17.003990}
}

@article{kouyoumjian1974,
  author  = {Kouyoumjian, Robert G. and Pathak, Prabhakar H.},
  title   = {A uniform geometrical theory of diffraction for an edge in a perfectly conducting surface},
  journal = {Proceedings of the IEEE},
  volume  = {62},
  number  = {11},
  pages   = {1448--1461},
  year    = {1974},
  doi     = {10.1109/PROC.1974.9651}
}

@article{deygout1966,
  author  = {Deygout, Jacques},
  title   = {Multiple knife-edge diffraction of microwaves},
  journal = {IEEE Transactions on Antennas and Propagation},
  volume  = {14},
  number  = {4},
  pages   = {480--489},
  year    = {1966},
  doi     = {10.1109/TAP.1966.1138719}
}

@article{bullington1947,
  author  = {Bullington, Kenneth},
  title   = {Radio propagation at frequencies above 30 megacycles},
  journal = {Proceedings of the IRE},
  volume  = {35},
  number  = {10},
  pages   = {1122--1136},
  year    = {1947},
  doi     = {10.1109/JRPROC.1947.232600}
}

@techreport{hufford1982,
  author      = {Hufford, G. A. and Longley, A. G. and Kissick, W. A.},
  title       = {A guide to the use of the {ITS} Irregular Terrain Model in the area prediction mode},
  institution = {U.S. Department of Commerce, National Telecommunications and Information Administration},
  address     = {Boulder, CO, USA},
  number      = {NTIA Report 82-100},
  year        = {1982},
  month       = apr
}

@techreport{3gpp36101,
  author      = {{3GPP}},
  title       = {Evolved universal terrestrial radio access ({E-UTRA}); user equipment ({UE}) radio transmission and reception},
  institution = {3rd Generation Partnership Project},
  type        = {Tech. Spec.},
  number      = {TS 36.101 V18.10.0 (Release 18)},
  month       = jul,
  year        = {2025}
}

@techreport{3gpp36104,
  author      = {{3GPP}},
  title       = {Evolved universal terrestrial radio access ({E-UTRA}); base station ({BS}) radio transmission and reception},
  institution = {3rd Generation Partnership Project},
  type        = {Tech. Spec.},
  number      = {TS 36.104 V18.6.0 (Release 18)},
  month       = jan,
  year        = {2025}
}

@techreport{sfcg32,
  author      = {{Space Frequency Coordination Group}},
  title       = {Communication and positioning, navigation, and timing frequency allocations and sharing in the lunar region},
  institution = {SFCG},
  number      = {Recommendation SFCG 32-2R5},
  month       = jun,
  year        = {2023}
}

@inproceedings{paszke2019,
  author    = {Paszke, Adam and Gross, Sam and Massa, Francisco and Lerer, Adam and Bradbury, James and Chanan, Gregory and Killeen, Trevor and Lin, Zeming and Gimelshein, Natalia and Antiga, Luca and Desmaison, Alban and K{\"o}pf, Andreas and Yang, Edward and DeVito, Zachary and Raison, Martin and Tejani, Alykhan and Chilamkurthy, Sasank and Steiner, Benoit and Fang, Lu and Bai, Junjie and Chintala, Soumith},
  title     = {{PyTorch}: An imperative style, high-performance deep learning library},
  booktitle = {Advances in Neural Information Processing Systems 32},
  address   = {Vancouver, BC, Canada},
  year      = {2019},
  pages     = {8024--8035}
}

@misc{neumann2010gdr,
  author       = {Neumann, G. A.},
  title        = {{Lunar Orbiter Laser Altimeter} Gridded Data Record, {LRO-L-LOLA-4-GDR-V1.0}},
  howpublished = {NASA Planetary Data System, Geosciences Node},
  year         = {2010},
  note         = {{LOLA} EXTRAS illumination and permanently shadowed region maps, \url{https://pds-geosciences.wustl.edu/missions/lro/lola.htm}}
}

@misc{creech2025hls,
  author       = {Creech, S.},
  title        = {Update on {NASA}'s {Human Landing System} ({HLS}) Program: Public-private partnership advancing {Artemis} sustainable lunar exploration},
  howpublished = {Presented at the 76th Int. Astronautical Congress, Sydney, Australia; NASA Marshall Space Flight Center, NTRS document 20250008728},
  year         = {2025},
  note         = {\url{https://ntrs.nasa.gov/citations/20250008728}}
}

@inproceedings{shumate2010,
  author    = {Shumate, Sid E.},
  title     = {{Longley-Rice} and {ITU-P.1546} combined: {A} new international terrain-specific propagation model},
  booktitle = {Proc. IEEE 72nd Veh. Technol. Conf. (VTC 2010-Fall)},
  address   = {Ottawa, ON, Canada},
  year      = {2010}
}

@misc{nasaim2end,
  author = {{NASA}},
  title  = {{NASA} Receives Some Data Before {Intuitive Machines} Ends Lunar Mission},
  howpublished = {Release 25-025, \url{https://www.nasa.gov/news-release/nasa-receives-some-data-before-intuitive-machines-ends-lunar-mission/}},
  year   = {2025},
  month  = mar
}

\end{document}